\documentclass[11pt,a4paper]{article}
\usepackage[T1]{fontenc}
\usepackage[numbers,sort&compress]{natbib}
\usepackage{jheppub}
\usepackage{orcidlink}
\usepackage{mathtools,mathrsfs,bm,booktabs,array,microtype,placeins,etoolbox}
\apptocmd{\thebibliography}{\setlength{\itemsep}{1.5pt plus .2ex minus .05ex}}{}{}
\usepackage{chngcntr}
\counterwithout{equation}{section}

\let\OriginalTableOfContents\tableofcontents
\renewcommand{\tableofcontents}{\begingroup\renewcommand{\baselinestretch}{0.94}\small\OriginalTableOfContents\endgroup}
\renewcommand{\afterTocRuleSpace}{\bigskip\bigskip\clearpage}
\allowdisplaybreaks[1]
\graphicspath{{figures/}{./}}
\newcommand{\dd}{\mathrm{d}}
\newcommand{\Nfive}{\mathcal N_5}

\newcommand{\tr}{\operatorname{tr}}
\newcommand{\HT}{\mathrm{HT}}

\newcommand{\GeV}{\mathrm{GeV}}
\newcommand{\fm}{\mathrm{fm}}
\newcommand{\PaperFigure}[4][tbp]{%
 \begin{figure}[#1]\centering
 \includegraphics[width=\linewidth,height=.73\textheight,keepaspectratio]{#2.pdf}%
 \caption{#3}\label{#4}\end{figure}}
\hypersetup{pdftitle={Glueball gravitational form factors in dynamical holography},pdfauthor={Kiminad A. Mamo}}
\title{Glueball gravitational form factors in dynamical holography}
\author{Kiminad A.~Mamo\,\orcidlink{0000-0003-3123-4398}\,}
\affiliation{Department of Physics, University of Connecticut,\\
Storrs, Connecticut 06269, USA}
\emailAdd{kiminad.mamo@uconn.edu}

\abstract{We present the first holographic study of gravitational form factors (GFFs) of scalar, pseudoscalar and canonical-spin-averaged tensor glueball targets. These observables test gravitational cubic couplings whose physical validity is not established by spectra and static-force calibration. We construct a dynamical Einstein--scalar--axion representation inspired by the Babington--Crooks--Evans adjoint-mass deformation of $\mathcal N=4$ super-Yang--Mills theory and its Type-IIB supergravity dual. Leading ultraviolet source data are retained, while a soft-wall-like infrared reconstruction incorporates asymptotically linear radial mass-squared trajectories. Spectra and Wilson observables fix the parameters at a common Sommer scale without GFF input. The full constrained cubic vertices fail to describe 27 correlated scalar lattice observations, giving $\chi^2/\mathrm{dof}=73.60/27=2.73$. We propose normalized heavy-target (HT) effective cubic couplings for pure-glue matching. HT adopts leading radial shapes suggested by infinitely heavy targets at fixed probe dynamics and normalizes their overlap weights on the physical ground states. With the background, modes and propagators unchanged, HT gives $\chi^2/\mathrm{dof}=16.98/27=0.63$, without numerical adjustment to GFF data. Tensor and scalar source overlaps determine $A$ and the physical normalized trace; the exact finite-mass energy--momentum-tensor relation determines $D$. We predict the pseudoscalar and tensor form factors and three-target Breit pressure and shear profiles. On the reference infrared-decaying axion branch, the HT Breit mass radii are $0.276$, $0.300$ and $0.311\,\mathrm{fm}$, respectively. The results identify gravitational cubic matching as a distinct step beyond background calibration and provide a prescription for further lattice tests.}
\begin{document}
\maketitle
\clearpage

\section{Introduction}
\label{sec:introduction}
We present the first holographic study of glueball gravitational form factors
(GFFs). These metric matrix elements probe cubic vertices coupling one source
to two normalizable glueball modes. Spectra constrain the quadratic dynamics,
while the static force probes an effective string on the same background.
Those calibration observables do not independently test the gravitational
cubic couplings. GFFs supply this additional interaction information for the
holographic dictionary of pure Yang--Mills theory.

The parent is the adjoint-mass deformation of $\mathcal N=4$
super-Yang--Mills theory introduced by Babington, Crooks and Evans
(BCE), with its Type-IIB supergravity description~\cite{bce,apreda}.
Equal adjoint-fermion masses and induced scalar masses motivate a
low-energy pure-glue sector. Parametric separation of all unwanted
matter thresholds from a fixed confinement scale lies beyond the
controlled classical-supergravity regime~\cite{polchinski_strassler};
the classical analysis found fermionic bound states comparable in mass
to glueballs~\cite{apreda}. We retain the leading BCE mass-source and
metric data and reconstruct a soft-wall-like dynamical infrared geometry
with an imposed asymptotically linear radial mass-squared spectrum. The
specified local action fixes the background, fluctuation operators and
full cubic vertices consistently.

\emph{Background calibration and gravitational-interaction matching are
separate steps.} Glueball spectra and Wilson observables determine a
single background at a common Sommer scale. No GFF or radius enters this
parameter determination. The resulting target modes and probe propagators
supply a common basis for comparing gravitational cubic couplings against
the scalar lattice GFFs and predicting the other target channels.
We propose a normalized \emph{heavy-target} (HT) prescription for the
effective elastic cubic couplings to be matched to the pure-glue theory.\footnote{%
One possible quantum-effective interpretation views the reconstruction as a
finite model for selected kernels of a reduced one-particle-irreducible
bulk functional motivated by the BCE system. Connected responses of an exact quantum
functional are assembled from trees of its exact propagators and proper
vertices~\cite[Sec.~21]{srednicki}. In this interpretation, the reconstructed
background and quadratic operators model the propagating sector, while
HT supplies a candidate elastic cubic response of the matched pure-glue
sector. Exactness belongs to the fully matched quantum functional; the
present reconstruction models the selected response kernels.}

The non-GFF calibration constrains the background and quadratic response
operators within the chosen model. The reconstructed action also imposes
relations among the cubic gravitational couplings that these inputs do not
independently test. Full denotes the vertices derived from that action,
including its potential derivatives, lapse and shift constraints,
scalar--trace source map and canonical particle normalization. HT instead
builds the elastic couplings from normalized target-mode densities and the
two metric-source propagators. It retains the calibrated radial dynamics
while replacing the detailed cubic matching by a common overlap
prescription. The GFF comparison tests this additional interaction
information.

Both assignments use the same background, target modes, probe propagators
and physical scale. Each HT overlap is identified with its physical
metric-source channel: the trace is a scalar-source response, and the
exact finite-mass EMT identity cancels tensor-exchange poles from it.
Normalization of the effective overlap couplings leaves the conserved
EMT operator and canonically normalized target states unchanged.

The large-target-eigenvalue organization of the full on-shell amplitudes,
with probe dynamics held fixed, selects the radial shapes used in HT.
We normalize these shapes on the physical ground states and adopt their
spacelike continuation as a prescription to be tested. This is a
phenomenological coupling postulate, rather than a controlled expansion
for the ground states. The heavy-target construction is distinct from
heavy-adjoint decoupling, which changes the parent theory along a
threshold-matching trajectory. We postulate that the normalized leading
radial couplings describe the pure-Yang--Mills gravitational response.
At the common spectra-and-Wilson-calibrated point, the full vertices fail
to describe the scalar lattice GFFs, whereas HT describes them and yields
predictions for the pseudoscalar and tensor channels.

This starting point differs from improved holographic QCD, whose
ultraviolet dilaton is organized around the Yang--Mills running coupling
and beta function~\cite{ihqcd1,ihqcd}. Related reconstructions combine
relevant ultraviolet deformations or backreacted graviton--dilaton
backgrounds with confining infrared behavior~\cite{ballonbayona2018,lihuang2013}.
Here the retained mass-deformation source and the HT cubic couplings
organize the gravitational-response matching.

Mode-density overlaps have precedents in holographic gravitational form
factors~\cite{bdt_gff,abidin_vector,fujita_dterm}. Tensor and scalar
gravitational channels determine nucleon form factors, radii and stresses
in Refs.~\cite{mamozahed2020,mamozahed2022}; top-down pion calculations
likewise exhibit glueball-mediated form factors and scalar-channel
contributions to pressure~\cite{fujii2024,fujii2025}.
Those form-factor calculations concern pion, nucleon or vector-meson
targets. Here the glueballs themselves are the external targets of the
metric matrix element. Earlier gauge-invariant studies developed
active-scalar/metric three-point functions and on-shell glueball
couplings~\cite{bianchi2003three,muckprisco2004,muck2005}.
The calibration uses continuum glueball spectra~\cite{at} and the quenched
static force~\cite{necco}. Scalar GFFs and their covariance~\cite{gff,release}
are reserved for the subsequent comparison.
Preliminary lattice GFFs were reported in Ref.~\cite{abbott2024}.
Wavefunction and constituent-gluon studies give complementary context
for compact scalar structure~\cite{gff,shuryakzahed2026}; their size
measures differ from the Breit energy radius used here.

The main text presents the fixed background, the Full and HT vertices,
and the resulting curves and radii. The appendices collect the
scalar--trace source map, the constrained cubic derivation, the leading
HT weights and positivity proof, and a concise account of the reference
calibration.

\subsection{Summary of main results}
\label{sec:results_summary}
At the common Sommer scale, write
\[
 \begin{gathered}
 \xi=mz,\qquad x=Q^2/m^2,\qquad \mu_X=M_X^2/m^2,\qquad X=S,a,T,\\
 P=e^{3A(\xi)},\qquad c=-\bar\phi'/A'.
 \end{gathered}
\]
The dynamical background determines the warp factor $A$, canonical scalar
$\bar\phi$, axion kinetic function $Z$, and normalized ground-state target
wavefunctions $u_X$. They obey
\[
 \int Pu_S^2\dd\xi=1,\qquad \int PZu_a^2\dd\xi=1,\qquad
 \int Pu_T^2\dd\xi=2.
\]
The last factor implements the tensor polarization normalization. Define
unit target densities
\[
 p_S=Pu_S^2,\qquad p_a=PZu_a^2,\qquad p_T=Pu_T^2/2.
\]
The tensor and scalar metric-source bulk-to-boundary propagators $H_T,H_S$
are shared by the three targets:
\[
 (PH_T')'=xPH_T,\qquad(Pc^2H_S')'=xPc^2H_S,\qquad H_Y(0,x)=1,
\]
with the regular infrared prescription. Their relation to independent
boundary sources is derived in Appendix~\ref{sec:sourcemap}.

The leading effective HT cubic couplings are encoded by two normalized
radial weights. Their contractions with the unchanged metric probes give
\[
 \boxed{\mathcal A_X(x)=\int\dd\xi\,p_XH_T,\qquad
 \mathcal S_X(x)=\frac{\int\dd\xi\,c^2p_XH_S}{\langle c^2\rangle_X},
 \qquad \langle c^2\rangle_X=\int\dd\xi\,c^2p_X.}
\]
The reference HT matching prescription defines the finite-mass effective
EMT through
\[
 \boxed{ A_{X,\HT}=\mathcal A_X,\qquad A_{X,\HT}^{\rm S}=\mathcal S_X,\qquad
 D_{X,\HT}=\frac{4\mu_X}{3x}(\mathcal S_X-\mathcal A_X)-\frac13\mathcal A_X. }
\]
Here $A_X^{\rm S}=\Theta_X/(2M_X^2)$ is the normalized physical trace.
The overlaps keep the tensor and scalar probe spectra in their respective
source channels. Both equal one at $x=0$; their differentiated integrals
determine $D(0)$. Tensor entries refer to the canonical-spin monopole,
denoted by bars in the channel-specific formulas.

All curves use the spectra-and-Wilson-calibrated parameters in
Section~\ref{sec:model}. Figure~\ref{fig:HTsummary} shows the HT predictions
and the scalar lattice measurements; Figure~\ref{fig:fullHT} compares Full
and HT directly. At fixed calibrated parameters their scalar GFF scores
are $\chi^2/\mathrm{dof}=2.73$ and $0.63$, respectively, for 27
correlated observations. The pseudoscalar entries use the infrared-decaying
axion branch specified in Section~\ref{sec:model}.

\PaperFigure[!htbp]{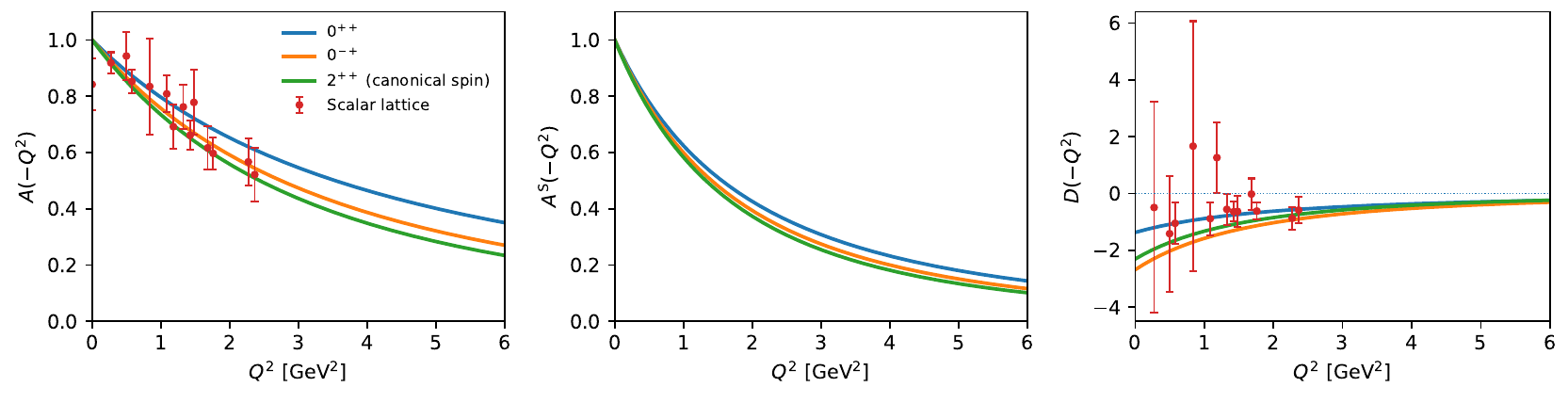}{HT ground-state predictions at the common
Sommer scale. Panels show $A_X=\mathcal A_X$, the physical
trace $A_X^{\rm S}=\mathcal S_X$, and $D_X$. The three targets use
the same dynamical background and probe operators, with the reference
axion branch in the pseudoscalar channel. Scalar lattice $A,D$ points~\cite{gff,release}
retain their published momentum axis and marginal errors; the full
covariance enters the comparison in Appendix~\ref{sec:calibration}.}{fig:HTsummary}

\begin{table}[tbp]\centering\small
\begin{tabular}{lrrr}\toprule
Target & $r_{\rm mass,BF}$ [fm] & $D(0)$ & Lattice radius [fm]\\\midrule
$0^{++}$ & 0.276 & -1.371 & $0.263(31)$\\
$0^{-+}$ & 0.300 & -2.685 & Not available\\
$2^{++}$ (canonical spin) & 0.311 & -2.308 & Not available\\\bottomrule
\end{tabular}
\caption{HT predictions at the reference point in Table~\ref{tab:mainparameters}.
The pseudoscalar result conditions on the reference infrared-decaying axion
branch. The scalar lattice radius uses the same Breit definition and is
quoted in its published units~\cite{gff}; its $n$-pole extrapolation includes
$D(0)$ rather than a measured forward point. No GFF or radius enters the
calibration. Table~\ref{tab:HTuncertainties} gives conditional parameter
errors; Appendix~\ref{sec:calibration} discusses branch and relative-scale
sensitivity.}
\label{tab:HTsummary}
\end{table}

Figure~\ref{fig:mainstressphysical} shows the pressure and shear obtained
by numerical Fourier transformation of the Full and HT form factors.
The transform retains each target's exact Breit recoil factor.
The definitions and physical interpretation are given in
Section~\ref{sec:stress}.
\PaperFigure{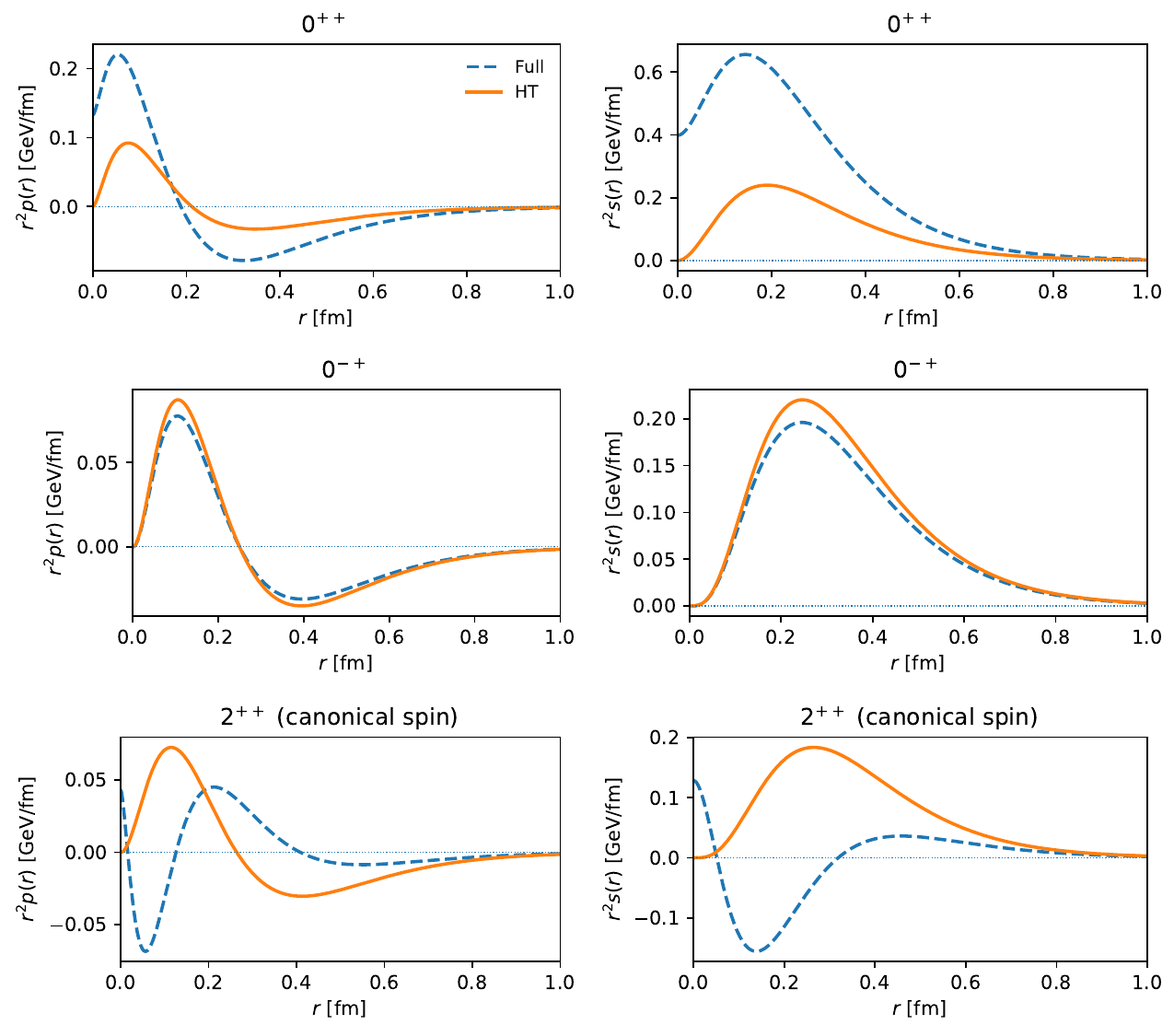}{Ground-state Breit pressures and shears at the
common Sommer scale. Dashed curves are the full dynamical vertices and solid
curves are the normalized HT model. Each target retains its own mass in
$E_Q=\sqrt{M_X^2+Q^2/4}$. Radial weights are $r^2p$ and $r^2s$, with no
additional $4\pi$. Pseudoscalar curves use the reference axion branch;
tensor curves refer to the canonical-spin monopole. The short-distance
structure follows from the specified all-transfer continuation of the
model form factors.}
{fig:mainstressphysical}
\FloatBarrier

\section{BCE-inspired dynamical reconstruction}
\label{sec:model}
\subsection{Background solution and fixed parameters}
The reconstruction preserves the leading mass-deformation source and
metric behavior inherited from the BCE parent while replacing its
infrared completion. The geometry, scalar potential and kinetic
functions form one dynamical system; an effective string insertion
completes the Wilson sector. The Lorentzian functional is
\begin{equation}
 S_5=\Nfive\int\dd^5x\sqrt{-g}
 \left[R-\frac12K(b)(\partial b)^2
 -\frac12Z(b)(\partial a)^2-V(b)\right]+S_{\partial}.
 \label{eq:mainaction}
\end{equation}
The single background used below is given explicitly in the field chart
$\bar b=mz$, $\bar a=0$ by
\begin{equation}
\begin{aligned}
 \dd s_E^2&=e^{2A_E(z)}
 (\dd z^2+\eta^{\rm bulk}_{\mu\nu}\dd x^\mu\dd x^\nu),\\
 e^{2A_E(z)}&=\frac{L^2}{z^2}\exp[-y_{\rm IR}(mz)^2]
 \left(1+\frac{(mz)^2}{b_c^2}\right)^{(y_{\rm IR}-y_{\rm UV})b_c^2},
 \qquad \bar b=mz,\quad\bar a=0 .
\end{aligned}
\label{eq:mainbackground}
\end{equation}
Here $\eta^{\rm bulk}=(-,+,+,+)$ and $0<z<\infty$.
The AdS radius $L$ fixes the bulk unit; the overall normalization
$\Nfive L^3$ cancels from the normalized form factors.
Equivalently, $A_E=\log(L/z)+F(b)$, with
\begin{equation}
 F(b)=-\frac{y_{\rm IR}}2b^2
 +\frac{(y_{\rm IR}-y_{\rm UV})b_c^2}{2}
       \log\left(1+\frac{b^2}{b_c^2}\right),\qquad y_{\rm UV}=\frac1{18}.
 \label{eq:warpansatz}
\end{equation}
All parameters are fixed in Table~\ref{tab:mainparameters}.

\begin{table}[!htbp]\centering\small
\begin{tabular}{lrlr}\toprule
Parameter & Fixed value & Parameter & Fixed value\\\midrule
$y_{\rm UV}$ & $1/18$ & $y_{\rm IR}$ & 0.839312707\\
$b_c$ & 9.563298310 & $b_a$ & 3\\
$\zeta_1$ & $-2.873993573$ & $\zeta_2$ & 3.520885950\\
$b_W$ & 3.011032523 & $\gamma$ & 0.187019507\\
$r_0$ [fm] & 0.49 & $m$ [GeV] & 1.813416168\\\bottomrule
\end{tabular}
\caption{Fixed background and effective-string parameters for the primary
Full and HT curves. The six fitted coefficients are
$y_{\rm IR},b_c,\zeta_1,\zeta_2,b_W,\gamma$; $y_{\rm UV}$ and $b_a$ are fixed
inputs. The mass scale follows from $m=(mr_0)\hbar c/r_0$, with
$mr_0=4.503053$ and $\hbar c=0.1973269804\,\GeV\,\fm$.
Digits specify the numerical parameter point, not independent parameter
precision. Predictions condition on these functional forms, $b_a=3$ and
the infrared-decaying axion branch $\zeta_1<0$.}
\label{tab:mainparameters}
\end{table}

The same warp input can be specified by its derivative profile $Y(b)$:
\begin{equation}
 F_b(b)=-bY(b),\qquad
 F(b)=-\int_0^b\dd\beta\,\beta Y(\beta),\qquad F(0)=0.
 \label{eq:warpDerivativeProfile}
\end{equation}
For the reference background,
\begin{equation}
 Y(b)=y_{\rm UV}+
       (y_{\rm IR}-y_{\rm UV})\frac{b^2}{b^2+b_c^2},\qquad
 Y_b=\frac{2b b_c^2(y_{\rm IR}-y_{\rm UV})}{(b^2+b_c^2)^2}.
 \label{eq:fixedprofileY}
\end{equation}
This profile gives Eq.~\eqref{eq:warpansatz} upon integration.\footnote{%
Replacing it by
$Y_{\exp}=y_{\rm UV}+(y_{\rm IR}-y_{\rm UV})[1-e^{-b^2/b_c^2}]$
and rematching its two warp coefficients to the same four scalar/tensor
masses at fixed Wilson sector and scale changes the HT forward $D$-terms
by less than $0.005\%$ and their radii by less than $0.03\%$.
The scalar comparison gives $\chi^2_{AD}/27=0.6289$ for the reference
profile and $0.6294$ for the exponential profile. Their scalar
large-transfer powers coincide, with leading coefficients differing by
less than $0.1\%$. See \hyperref[app:warpInterpolation]{``Focused
parametrization checks''} in Appendix~\ref{sec:calibration} for the
closed-form $F_{\exp}$ and the change of asymptotic radial slope.}
Using $F_b=-bY$ and $F_{bb}=-Y-bY_b$, the Einstein equations give the
scalar kinetic function and potential:
\begin{align}
 K(b)&=6\big[b^2Y(b)^2+3Y(b)+b Y_b(b)\big],
 \label{eq:kineticreconstruction}\\
 L^2V(b)&=-3e^{-2F(b)}
 \big[4+5b^2Y(b)-b^3Y_b(b)+3b^4Y(b)^2\big].
 \label{eq:mainreconstruction}
\end{align}
These expressions are equivalent to
$K=6(F_b^2-F_{bb}-2F_b/b)$ and
$L^2V=-3e^{-2F}(4-6bF_b+b^2F_{bb}+3b^2F_b^2)$.
For $b_a=3$, the reference infrared-decaying axion branch has
\begin{equation}
 Z(b)=\left(1+\frac{b^2}{9}\right)^{-2.873993573}
       \exp\left(\frac{3.520885950\,b^2}{b^2+9}\right).
 \label{eq:mainansatz}
\end{equation}
In this family $Z\sim e^{\zeta_2}(b/b_a)^{2\zeta_1}$ at large $b$.
We select the solution with $\zeta_1<0$ as part of the reference model.
The same two pseudoscalar masses also admit infrared-growing solutions;
positivity of $Z$ alone does not select the branch. Section~\ref{sec:unfittedtests}
describes their form-factor implications.
Thus the geometry and the functions $K,V,Z$ are explicit at the chosen
parameter point. The canonical scalar and the scalar--metric mixing
function are determined by
\begin{equation}
 \phi(b)=\int_0^b\sqrt{K(\beta)}\,\dd\beta,\qquad
 c(b)=\frac{b\sqrt{K(b)}}{1+b^2Y(b)}.
 \label{eq:canonicalprofile}
\end{equation}
The first relation is a field redefinition given by quadrature; the
background is analytic in the chart $b=mz$. Cubic potential derivatives
are taken with respect to $\phi$, as in Appendix~\ref{app:radial}.
Since $Y>0$ and $Y_b>0$ for the fixed parameters, $K$ is positive;
$Z$ is also positive. The scalar equation follows from the Einstein
equations on this monotone trajectory.

The leading ultraviolet behavior is
\begin{equation}
 \bar\phi=mz+O(z^3),\qquad
 A_E=\log(L/z)-m^2z^2/36+O(z^4),
 \label{eq:parentuv}
\end{equation}
which matches the BCE source data~\cite{bce,apreda}. The reconstructed
canonical potential has
\begin{equation}
 L^2V(\phi)=-12-\frac32\phi^2-\frac1{12}\phi^4+O(\phi^6).
 \label{eq:quarticcomparison}
\end{equation}
The classical BCE restriction has the same quadratic term but the
quartic coefficient $-1/8$ rather than $-1/12$~\cite{bce,apreda}.
The dimension-three source and its leading backreaction are shared.
Within the chosen analytic $F(b^2)$ family, the field equations fix the
reconstructed quartic: for $L^2V=-12-3\phi^2/2+v_4\phi^4+\cdots$,
the scalar equation gives
\begin{equation}
 \bar\phi(z)=mz+z^3\left[\phi_{(3)}+
 m^3\left(2v_4+\frac16\right)\log\frac z\ell\right]+\cdots .
 \label{eq:quarticlog}
\end{equation}
Here $\ell$ is a reference length. Thus $v_4=-1/12$ removes the
source-induced logarithm, whereas the BCE
value $-1/8$ gives $-m^3z^3\log(z/\ell)/12$. Higher interactions follow
from the reconstruction. Appendix~\ref{sec:sourcemap} fixes the physical
trace source.

The six parameters in Table~\ref{tab:mainparameters} minimize a common
spectra-and-Wilson objective: two masses per target channel, ten
static-force values, and $r_0\sqrt\sigma$. The Sommer condition
$r_0^2F_{Q\bar Q}(r_0)=1.65$ and $r_0=0.49\,\fm$ fix the dimensional
scale. The 17 calibration observations and six parameters give
$\chi^2_{\rm cal}/\mathrm{dof}=26.83/11=2.44$.
All Full and HT curves use this same point, with GFFs excluded from
parameter determination. Figure~\ref{fig:spectraforce} shows the spectral
and force comparisons; Appendix~\ref{sec:calibration} gives the objective
and the propagated parameter uncertainties.
\FloatBarrier

\subsection{Quadratic responses and the infrared spectrum}
\label{sec:ir}
Let $P=e^{3A_E}$ and $c=-\bar\phi'/A_E'$, with radial primes here denoting
$\dd/\dd z$. The tensor, active-scalar and axion weights are
\begin{equation}
 w_T=P,\qquad w_S=Pc^2,\qquad w_a=PZ.
 \label{eq:mainweights}
\end{equation}
For the active gauge-invariant scalar--metric fluctuation $q$, the
minimally weighted variable is $v=q/c$. Its normalizable tower couples
to the independent scalar source $j_\phi$ and the metric trace source
$j_\sigma$ through
\begin{equation}
 J_q\equiv\lim_{z\to0}\frac{q(z)}{z}=j_\phi+m j_\sigma .
 \label{eq:mainsourcemap}
\end{equation}
The ultraviolet index $\Delta=3$ belongs to the inherited parent
mass-deformation operator $\mathcal O_\phi$. Varying $j_\sigma$ at fixed
$j_\phi$ generates the constrained trace response, including the
dimension-four source-weighted contribution $m\mathcal O_\phi$ and the
associated source terms derived in Appendix~\ref{sec:sourcemap}.
The coupled scalar--metric tower supplies our finite-scale representation
of the Yang--Mills trace channel.\footnote{%
For canonically normalized gauge fields, the vacuum-subtracted
pure-Yang--Mills trace
$T^\mu{}_{\mu}=(\beta(g)/2g)[F^a_{\rho\sigma}F^{a\,\rho\sigma}]_R$
is renormalization-group invariant in physical matrix
elements~\cite{deldebbio_zwicky_trace}. The inherited ultraviolet mode
power characterizes the specified holographic continuation.}
The corresponding mode problems are
\begin{equation}
 -(w_Iv_{I,n}')'=M_{I,n}^2w_Iv_{I,n},
 \qquad
 \psi_{I,n}=\sqrt{w_I}\,v_{I,n},
 \qquad U_I=\frac{(\sqrt{w_I})''}{\sqrt{w_I}}.
 \label{eq:mainSL}
\end{equation}
The ultraviolet branch and infrared domain complete each operator.
The source solutions have the same domain prescription and a prescribed
nonnormalizable boundary coefficient. Their relation to the canonical
particle normalization is recorded in Appendix~\ref{app:radial}.

The selected classical mass-deformed branch ends at finite conformal
distance~\cite{bce,apreda}. For its normalizable radial problem, the
large-level eigenvalue law on an interval of length $\ell_z$ is
$M_n^2\sim(\pi n/\ell_z)^2$. The infrared reconstruction replaces
that finite-interval problem by a half-line oscillator asymptotic.
For this background,
\begin{equation}
\begin{aligned}
 F(b)&=-\frac12y_{\rm IR}b^2+O(\log b),\qquad c^2\to6,
 \qquad\log Z=O(\log b),\\
 \frac{U_I(z)}{m^2}&=\frac94y_{\rm IR}^2(mz)^2+O(1),\qquad
 \frac{M_{I,n}^2}{m^2}=6y_{\rm IR}n+O(1)\quad(n\to\infty).
\end{aligned}
\label{eq:mainreggeslope}
\end{equation}
The common asymptotic radial mass-squared slope
$s_\infty\equiv\lim_{n\to\infty}M_{I,n}^2/n=6y_{\rm IR}m^2$
is built into the reconstructed quadratic infrared geometry~\cite{ihqcd}. Here $n$
labels radial excitation at fixed channel $I$. The low-lying lattice
masses constrain the transition region and channel weights, while
$s_\infty$ characterizes the ultimate infrared oscillator limit.
At the calibrated point $s_\infty=16.56\,\GeV^2$; it is distinct from
the spacings of the first measured levels. The mode problem uses the
normalizable half-line domain.

\subsection{The effective Wilson insertion}
\label{sec:mainwilson}
The effective Wilson observable is defined by
\begin{equation}
 S_W=T_W\int\dd^2\sigma\,e^{2\Omega(b)}
          \sqrt{-\det P[g_{5,E}]},\qquad A_s=A_E+\Omega .
 \label{eq:mainstringaction}
\end{equation}
\begin{equation}
\begin{aligned}
 G_W(b)&=\frac7{72}b_W^2\log(1+b^2/b_W^2),
 \qquad \Omega=G_W-F,\\
 h(b)&=\frac{e^{2G_W(b)}}{b^2}
      =\frac{(1+b^2/b_W^2)^{p_W}}{b^2},
 \qquad p_W=\frac7{36}b_W^2.
\end{aligned}
 \label{eq:mainwilsonfamily}
\end{equation}
The leading dressing is $\Omega=b^2/8+\cdots$, motivated by the
source expansion of Ref.~\cite{apreda}. For $p_W>1$ there is a positive string wall at
$b_{\min}^2=b_W^2/(p_W-1)$. A positive minimum of the string factor gives
a confining classical area-action response~\cite{kinar}.
Here $G_W$ has zero explicit quadratic infrared term, even though $A_E$
has quadratic infrared asymptotics. The two requirements concern
different frame combinations.

For $\rho=mR$, turning point $b_*$ and $\gamma=T_WL^2$, the string integrals are
\begin{align}
 \rho(b_*)&=2\int_0^{b_*}\frac{\dd b}
              {\sqrt{h(b)^2/h(b_*)^2-1}},
 \label{eq:wilsonlength}\\
 \mathcal E(b_*)&=2\gamma\left[
  \int_0^{b_*}\left(\frac{h(b)^2}{\sqrt{h(b)^2-h(b_*)^2}}
                            -\frac1{b^2}\right)\dd b-\frac1{b_*}\right]
  +C_W,
 \label{eq:wilsonenergy}\\
 F_{Q\bar Q}(R)&=m^2\gamma h(b_*),\qquad
 \sigma=m^2\gamma h(b_{\min}).
 \label{eq:mainwilsonobs}
\end{align}
$\mathcal E=E/m$ and $E(r_0)=0$ fixes the additive energy convention. The subtraction removes the ultraviolet perimeter
divergence inside the integral. With $\Omega=G_W-F$, the string-frame
warp factor is $A_s=\log(L/z)+G_W$: the force depends on $G_W$ and its
normalization, rather than directly on $F$. The Einstein-frame and
string profiles share the radial coordinate and dimensional scale.
The reference predicts $r_0\sqrt\sigma=1.18196$, compared with
$1.160(6)$; the discrepancy of $3.7$ quoted input standard errors belongs
to the string sector and is unchanged by the Full--HT vertex comparison.

\FloatBarrier

Figure~\ref{fig:spectraforce} compares the resulting spectra and static
force with the lattice inputs. Full and HT have identical predictions
in these sectors because their quadratic dynamics and string insertion
are unchanged.
\PaperFigure{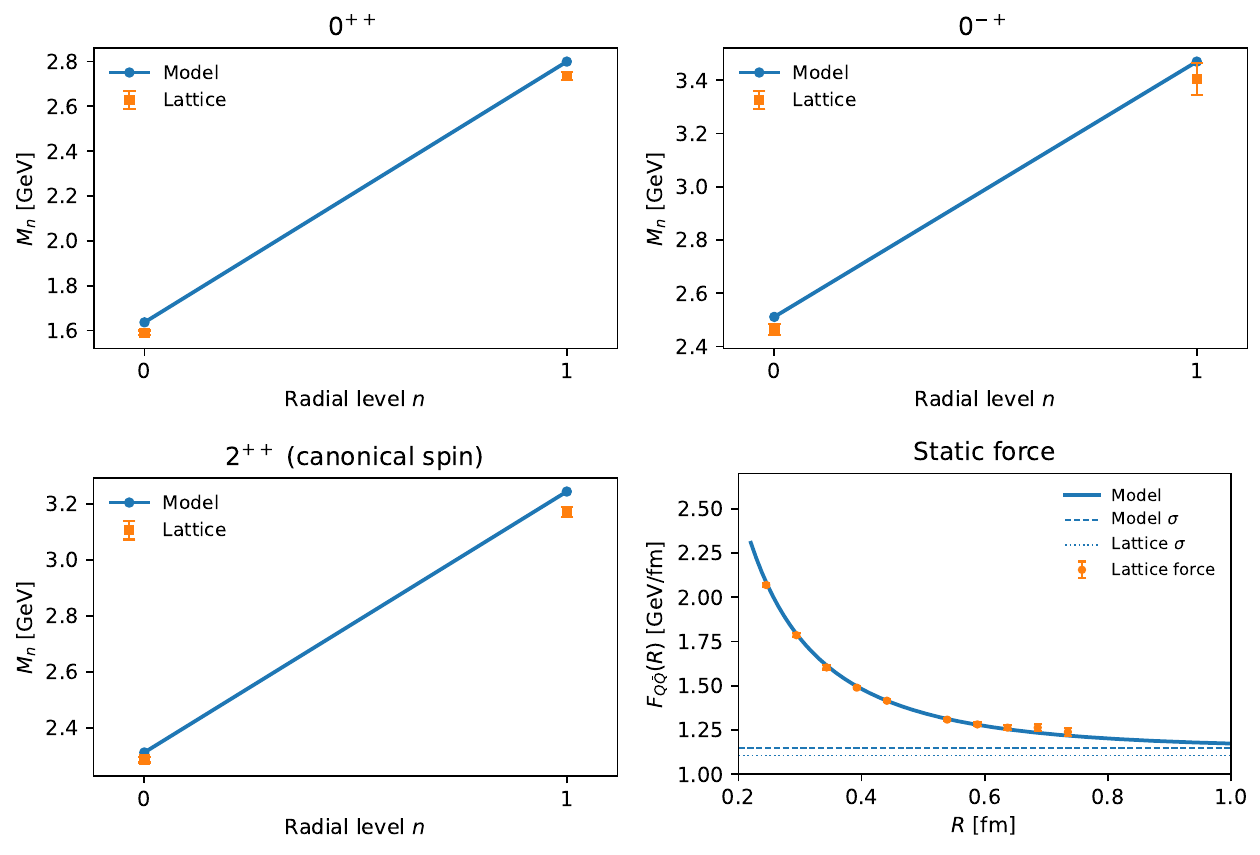}{Low-lying dynamical spectra and static force at
$r_0=0.49\,\fm$. The spectral lattice inputs are converted with their input
$r_0\sqrt\sigma=1.160(6)$ and the common Sommer convention. Displayed
spectral errors are the supplied pointwise errors. The force panel retains
the difference between the model and input string tensions. Changing the
cubic GFF prescription leaves these observables unchanged.}{fig:spectraforce}
\FloatBarrier

\section{Full cubic vertices and HT matching}
\label{sec:gffs}
The full gravitational cubic vertices follow from the reconstructed action
of Section~\ref{sec:model}. Their leading on-shell radial structure
motivates the effective cubic couplings adopted in HT. The derivation
below first gives the complete elastic vertex contractions, then defines
the normalized HT weights and their finite-mass EMT map. The background,
target modes and probe propagation remain common to both; lattice and
cross-channel comparisons test the proposed pure-glue coupling assignment.

\subsection{Metric sources, particle poles and EMT conventions}
A target matrix element is the double-pole coefficient of the source
three-point function. For spin zero,
\begin{equation}
 \langle O_X(p')T^{\mu\nu}(\Delta)O_X(-p)\rangle_c
 \sim \frac{f_X^2\mathcal M_X^{\mu\nu}(p',p)}
 {(p'^2-M_X^2)(p^2-M_X^2)}.
 \label{eq:mainpoles}
\end{equation}
Canonical particle normalization removes $f_X^2$. The full baseline is
obtained from Eq.~\eqref{eq:mainaction}, including its metric constraints
and variational boundary terms before the residues are extracted.
Appendices~\ref{sec:sourcemap} and \ref{sec:ESAADM} give this calculation.
With $\eta^{\rm phys}=(+,-,-,-)=-\eta^{\rm bulk}$,
$\Delta=p'-p$, $\bar p=(p'+p)/2$ and $t=-Q^2$,
\begin{equation}
 \mathcal M_X^{\mu\nu}=2\bar p^\mu\bar p^\nu A_X(t)
 +\frac12(\Delta^\mu\Delta^\nu-\eta^{\mu\nu}_{\rm phys}t)D_X(t).
 \label{eq:mainEMT}
\end{equation}
Writing $\Theta_X=2M_X^2A_X^{\rm S}$ gives
\begin{equation}
 D_X(-Q^2)=\frac{4M_X^2}{3Q^2}(A_X^{\rm S}-A_X)-\frac13A_X.
 \label{eq:maintrace}
\end{equation}
The superscript ${\rm S}$ labels the normalized metric trace. For the
tensor target we average the five diagonal equal-rest-spin matrix elements
after rotationless boosts. Bars denote this canonical-spin monopole,
whose conversion from the polarized tensor matrix element is specified in
Appendix~\ref{sec:polarizedtensor}~\cite{cotogno_multipoles}.

\subsection{Elastic contractions of the full cubic vertices}
On the background, the field-chart value is $\bar b=\xi=mz$. We use
$x=Q^2/m^2$, $\mu_X=M_X^2/m^2$, and the dimensionless warp factor
\[
 A(\xi)=A_E(\xi/m)-\log(Lm)=-\log\xi+F(\xi),\qquad P=e^{3A(\xi)}.
\]
Radial primes in the overlap formulas denote $\dd/\dd\xi$. The target
modes and metric-source profiles solve the operators specified in
Section~\ref{sec:ir}, with particle normalization as in
Appendix~\ref{app:radial}.
The full scalar response, with $N=-cu_S/6$ and
$B=-(cu_S'-c'u_S)/(6\mu_S)$, is
\begin{equation}
\begin{aligned}
 A_{S,\rm full}&=\int\dd\xi\,P[u_S^2H_T-2NBH_T'+xB^2H_T],\\
 \widehat\Theta_{S,\rm full}&=-\int\dd\xi\,P
 [(F_0+xF_1+x^2F_2)H_S+(G_0+xG_1)H_S'].
\end{aligned}
\label{eq:mainscalarfull}
\end{equation}
The $F_i,G_i$ are the complete canonical cubic coefficients in
Eq.~\eqref{app:scalarpolynomials}, including potential derivatives and
constraint contributions. The pseudoscalar response is
\begin{equation}
\begin{aligned}
 A_{a,\rm full}&=\int\dd\xi\,PZu_a^2H_T,\\
 \widehat\Theta_{a,\rm full}&=-\int\dd\xi\,P
 [(\mathcal E_0+x\mathcal E_1)H_S+\mathcal J_0H_S'].
\end{aligned}
\label{eq:mainaxionkernels}
\end{equation}
The coefficients are given in Eq.~\eqref{app:axionkernels}.
For the tensor define
\begin{equation}
 \mathcal F_T=\frac12\int\dd\xi\,Pu_T^2H_T,\qquad
 \mathcal G_T=\frac1{24\mu_T}\int\dd\xi\,Pc^2
 (\mu_Tu_T^2+u_T'^2)H_S.
\label{eq:maintensoroverlaps}
\end{equation}
The complete canonical-spin average is
\begin{equation}
 \overline A_{T,\rm full}=(1-2y/3+y^2/30)\mathcal F_T,\qquad
 \overline A^{\rm S}_{T,\rm full}=(1+y/3+y^2/30)\mathcal G_T,
 \qquad y=x/\mu_T.
\label{eq:maintensorfull}
\end{equation}
Appendix~\ref{sec:polarizedtensor} derives these expressions from the
Bose-summed polarized vertices and gives the unreduced radial kernels.
Radial identities and boundary terms are included before the full
response is organized asymptotically. Equivalent full-action
representatives yield the same on-shell functions.

\subsection{Leading radial structure at fixed probe dynamics}
\label{sec:HTdefinition}
This large-target organization identifies the leading radial vertex
structure used to define the HT effective couplings. The target eigenvalue is varied with the
background and probe operators held fixed. The target modes themselves remain on shell. Write their
canonical equations as
\begin{equation}
 -(W_Xu_X')'+W_XU_Xu_X=\mu_XW_Xu_X,\qquad
 (W_X,U_X)=\begin{cases}
 (P,c''/c+3A'c'/c),&X=S,\\
 (PZ,0),&X=a,\\
 (P,0),&X=T.
 \end{cases}
\label{eq:HTtargetequations}
\end{equation}
For a smooth radial factor $f$ and the stated vanishing endpoints,
\begin{equation}
 \int W_Xfu_X'^2\dd\xi
 =\mu_X\int W_Xfu_X^2\dd\xi
  -\int W_XfU_Xu_X^2\dd\xi
  +\frac12\int(W_Xf')'u_X^2\dd\xi.
\label{eq:HTmodeidentity}
\end{equation}
This is an integrated identity, not a pointwise replacement of target
derivatives. In the fixed-probe large-eigenvalue organization, it combines
the target radial and boundary kinetic contributions. The common leading
weights extracted from Eqs.~\eqref{eq:mainscalarfull}--\eqref{eq:maintensorfull} are
\begin{equation}
 \widetilde\rho^{(0)}_{A,X}=p_X,\qquad
 \widetilde\rho^{(0)}_{S,X}=\frac{c^2}{6}p_X,\qquad
 (p_S,p_a,p_T)=(Pu_S^2,PZu_a^2,Pu_T^2/2).
\label{eq:HTbareweights}
\end{equation}
For the pseudoscalar, the explicit $Z_{,\phi}$ contributions cancel between
the two leading kinetic terms. For the tensor, retaining all of
$\mathcal G_T$ would also retain a finite-target radial correction.
Appendix~\ref{app:HTderivation} gives the three-channel extraction.
The HT postulate retains this leading on-shell structure, normalizes it
on each physical ground state, and assigns its spacelike continuation.

\subsection{Normalized finite-mass HT model}
The bare scalar-channel overlap at zero transfer is
$\widetilde{\mathcal S}_X(0)=\langle c^2\rangle_X/6$. On the physical
low-lying states this differs from one (Table~\ref{tab:HTnormalizations}).
We preserve the leading radial shape and normalize its integral:
\begin{equation}
 \rho^{(0)}_{A,X}=p_X,\qquad
 \rho^{(0)}_{S,X}=\frac{c^2p_X}{\langle c^2\rangle_X},\qquad
 \langle c^2\rangle_X=\int c^2p_X\dd\xi.
\label{eq:HTweights}
\end{equation}
These weights specify the leading effective HT cubic couplings after
normalization of the two external target modes. The multiplicative
normalization defines the reference prescription; its denominator is
calculated separately for each mode, while the common source $H_S$ remains
unchanged. No additional coefficient is fitted.
Contracting the weights with the probe profiles gives
\begin{equation}
 \boxed{\mathcal A_X(x)=\int\rho^{(0)}_{A,X}H_T\dd\xi,\qquad
 \mathcal S_X(x)=\int\rho^{(0)}_{S,X}H_S\dd\xi.}
\label{eq:HToverlaps}
\end{equation}
We assign these overlaps directly to the tensor and scalar metric-source
responses:
\begin{equation}
 \boxed{A_{X,\HT}(x)=\mathcal A_X(x),\qquad
 A^{\rm S}_{X,\HT}(x)=\mathcal S_X(x).}
 \label{eq:HTtracecompletion}
\end{equation}
The exact finite-mass decomposition, Eq.~\eqref{eq:maintrace}, determines
\begin{equation}
 \boxed{D_{X,\HT}(x)=\frac{4\mu_X}{3x}
 [\mathcal S_X(x)-\mathcal A_X(x)]-\frac13\mathcal A_X(x).}
 \label{eq:HTmodel}
\end{equation}
The heavy-target organization selects the radial coupling shapes. The
finite-mass EMT conversion is kept exact, including the last term in
Eq.~\eqref{eq:HTmodel}. The hypothesis assigns this normalized total-EMT
response to pure Yang--Mills; it is tested over the lattice interval and
continued as defined for the spatial transforms.

\paragraph{Source-channel spectral selection.}
A one-particle tensor state cannot couple to the trace, because
$\langle0|T^{\mu\nu}|2,n\rangle=f_{2,n}\epsilon^{\mu\nu}$ and
$\eta_{\mu\nu}\epsilon^{\mu\nu}=0$.
Thus an isolated tensor-source pole at $t=t_{2,n}$ in $A$ must cancel
from $\Theta$. The EMT identity requires
\begin{equation}
 \operatorname*{Res}_{t=t_{2,n}}D(t)
 =\frac{4M_X^2-t_{2,n}}{3t_{2,n}}
   \operatorname*{Res}_{t=t_{2,n}}A(t).
 \label{eq:polecancellation}
\end{equation}
The assignment $A^{\rm S}=\mathcal S$ implements this cancellation:
$\mathcal S$ propagates through $H_S$ alone. For spin-zero targets the
separately conserved tensor and scalar channels may be written as
\begin{equation}
 \mathcal M^{\mu\nu}=
 2\left(\bar p^\mu\bar p^\nu-\frac{\bar p^2}{3}\Pi^{\mu\nu}\right)
 \mathcal A+
 \frac{2M_X^2}{3}\Pi^{\mu\nu}\mathcal S,\qquad
 \Pi^{\mu\nu}=\eta^{\mu\nu}_{\rm phys}
 -\frac{\Delta^\mu\Delta^\nu}{t}.
 \label{eq:irreducibleHT}
\end{equation}
The first term is traceless and the second has trace $2M_X^2\mathcal S$;
this is the conserved scalar/tensor decomposition~\cite{broniowski2025}.
The assignment of tensor and scalar exchange to these source channels
also underlies meson-dominance descriptions of pion GFFs~\cite{broniowski2024,ruizarriola2024}.
The equal forward overlaps remove the apparent kinematic singularity.
The tensor target uses the canonical-spin projection specified in
Appendix~\ref{sec:polarizedtensor}; its scalar trace has the same
one-particle selection rule. This does not determine the additional
polarized multipoles.

\begin{table}[tbp]\centering\small
\begin{tabular}{lrr}\toprule
Target & $\langle c^2\rangle_X$ & $\widetilde{\mathcal S}_X(0)=\langle c^2\rangle_X/6$\\\midrule
$0^{++}$ & 5.342939 & 0.890490\\
$0^{-+}$ & 6.395441 & 1.065907\\
$2^{++}$ (canonical spin) & 6.845852 & 1.140975\\\bottomrule
\end{tabular}
\caption{Bare leading scalar-channel normalizations on the physical ground states. Division by the final column is the finite-mass normalization-preserving completion in the HT definition. It preserves the retained radial shape and introduces no fit parameter.}
\label{tab:HTnormalizations}
\end{table}

\subsection{Forward normalizations and stress coefficient}
\label{sec:forwardmain}
The normalized weights and constant forward sources give
\begin{equation}
 \mathcal A_X(0)=\mathcal S_X(0)=A^{\rm S}_{X,\HT}(0)=1.
\label{eq:HTward}
\end{equation}
For $H_Y=1+xh_Y+O(x^2)$, the differentiated source is
\begin{equation}
 h_Y'(\xi)=-\frac1{w_Y(\xi)}\int_\xi^\infty w_Y(v)\dd v,
 \qquad h_Y(0)=0,\qquad(w_T,w_S)=(P,Pc^2).
\label{eq:maindifferentiatedsource}
\end{equation}
Hence the regular HT forward stress coefficient is
\begin{equation}
 D_{X,\HT}(0)=\frac{4\mu_X}{3}
 \left[\int\rho^{(0)}_{S,X}h_S\dd\xi-\int p_Xh_T\dd\xi\right]-\frac13.
\label{eq:HTforward}
\end{equation}
The baseline uses the complete source response in Eq.~\eqref{eq:maintrace}.
Terms of order $x/\mu_X$ in the two independent responses contribute at
order one to $D_X(0)$. Accordingly, the HT stress coefficient is a
finite-mass output of the HT prescription. The equal forward
normalizations ensure a regular quotient.

The HT effective cubic vertices here specify the coupling of a metric
source to two on-shell, canonically normalized target modes through
Eqs.~\eqref{eq:HTweights}--\eqref{eq:HTmodel}. These on-shell source--target couplings and their finite-mass EMT
map define the elastic model considered here.

\subsection{Positive spacelike overlaps}
\label{sec:HTpositivity}
\paragraph{HT positivity lemma.}
Suppose $w_T=P$ and $w_S=Pc^2$ are positive in the radial interior, the
unit ultraviolet source and regular infrared condition define a
positivity-preserving radial problem, and the target weights in
Eq.~\eqref{eq:HTweights} are integrable. Then, for finite $x\geq0$,
\begin{equation}
 0<\mathcal A_X(x)\leq1,\qquad
 0<\mathcal S_X(x)\leq1,\qquad
 \partial_x\mathcal A_X\leq0,\qquad\partial_x\mathcal S_X\leq0.
 \label{eq:HTpositivity}
\end{equation}
More strongly, on $x>0$ these two overlaps are completely monotone:
\begin{equation}
 (-1)^n\partial_x^n\mathcal A_X(x)\geq0,\qquad
 (-1)^n\partial_x^n\mathcal S_X(x)\geq0,
 \qquad n=0,1,2,\ldots .
 \label{eq:HTcompleteMonotonicity}
\end{equation}
The same inequalities extend to $x=0$ for every finite derivative there.
The positive radial resolvent proves the result, as shown in
Appendix~\ref{app:HTpositivityProof}. The fixed background satisfies these source and weight assumptions.

Because $A_{X,\HT}=\mathcal A_X$ and
$A^{\rm S}_{X,\HT}=\mathcal S_X$, the lemma applies directly to both
physical source-channel form factors. It imposes no sign on $D_X$ or on
local Breit energy and stress distributions.

\section{Form factors and mechanical distributions}
\label{sec:results}\label{sec:tests}
\subsection{Three-target responses and radii}
Figure~\ref{fig:fullHT} compares form factors obtained from the full
reconstructed cubic vertices and the leading effective HT cubic couplings
at the same spectra-and-Wilson-calibrated parameters.
Figure~\ref{fig:HTsummary} shows the three physical HT form factors on
common axes. The scalar $A,D$ observations test the predictions; the
physical traces and the other target channels have no GFF input. The pseudoscalar tensor-source overlap is identical in the
full and HT models. The explicit $Z_{,\phi}$ terms enter its full trace;
the HT scalar overlap retains the dependence through $PZu_a^2$ and $c^2$.
\PaperFigure{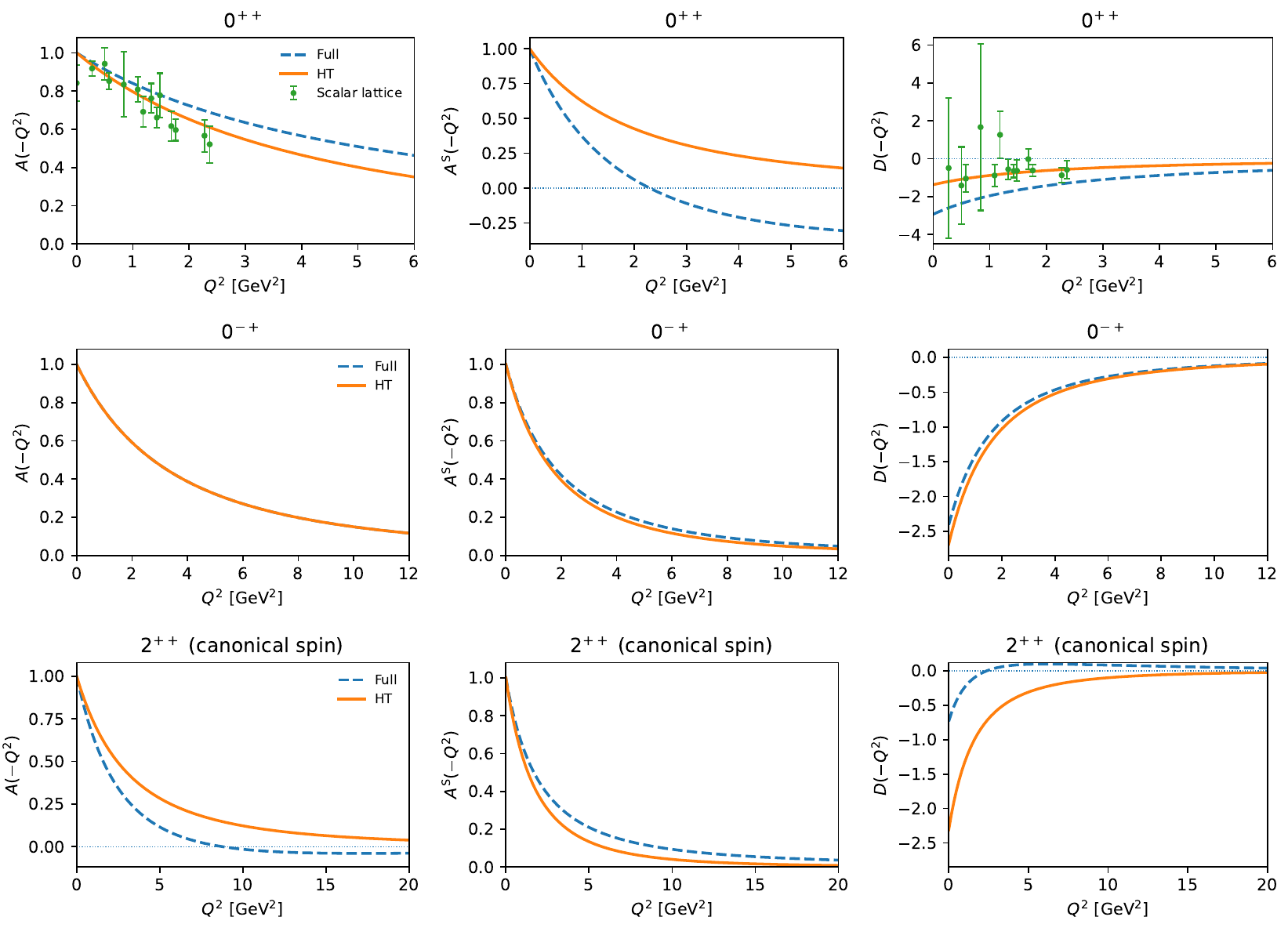}{Form factors from the full dynamical cubic vertices
(dashed) and the leading effective HT cubic couplings (solid). Rows show scalar, pseudoscalar and canonical-spin tensor
targets; columns show $A$, the physical normalized trace $A^{\rm S}$, and $D$.
The HT physical trace is the normalized $H_S$ overlap in
Eq.~\eqref{eq:HTtracecompletion}. Scalar points and marginal errors are
from Refs.~\cite{gff,release}. All curves use the same background and
physical scale, with the reference axion branch for the pseudoscalar.
Scalar points retain the published lattice momentum axis. No GFF
observations enter the calibration.}
{fig:fullHT}

The Full scalar trace changes sign near $Q^2=2.30\,\GeV^2$ on the
reference background, whereas the HT trace is positive by construction.
Positivity of a spacelike trace form factor is not a general physical
requirement.

{\clubpenalty=10000
At the fixed spectra-and-Wilson-calibrated parameter point, the published
$27\times27$ covariance gives $\chi^2_{AD}/\nu_{AD}=73.60/27=2.73$ for the
full vertices and $16.98/27=0.63$ for HT, with $\nu_{AD}=27$.
The HT form was developed with the scalar measurements available.
Its weights and overlap normalizations follow the stated prescription;
the numerical background parameters are determined from spectra and
Wilson observables without GFF input.
Appendix~\ref{sec:calibration} gives the statistical context and
relative-scale sensitivity.\par}

For context, the best of the three conventional $n$-pole benchmarks in
Appendix~\ref{sec:calibration} is a dipole fit with $\chi^2_{AD}=12.67$
after adjusting three GFF parameters using the same covariance. The HT
prediction lies $\Delta\chi^2=4.31$ above this benchmark without any GFF
parameter adjustment, while the full-vertex prediction lies
$\Delta\chi^2=60.93$ above it.

For the exact-recoil Breit energy kernel,
\begin{equation}
 \mathcal E_X(Q^2)=E_QA_X(-Q^2)+\frac{Q^2}{4E_Q}D_X(-Q^2),
 \qquad E_Q=\sqrt{M_X^2+Q^2/4},
\label{eq:energykernel}
\end{equation}
the invariant $A$-slope radius and the Breit energy radius are
\begin{equation}
 r_{A,X}^2=-\frac{6(\hbar c)^2}{m^2}\dot A_X(0),\qquad
 r_{{\rm mass,BF},X}^2=r_{A,X}^2-
 \frac{3(\hbar c)^2}{4M_X^2}[1+2D_X(0)].
\label{eq:mainmassradius}
\end{equation}
Dots denote derivatives in $x$. The second definition is the Breit
energy-radius definition in Eq.~(9) of Ref.~\cite{gff}. The scalar HT
prediction, $0.276\,\fm$, agrees with the lattice result
$0.263(31)\,\fm$ in that Breit definition, retaining the published lattice
length convention. The latter uses an $n$-pole extrapolation of the
measured $A,D$ functions. The possible relative length conversion is
quantified in Appendix~\ref{app:scaleMatching}. The Full radius is
$0.312\,\fm$. On the reference axion branch, the pseudoscalar HT mass
radius is $0.300\,\fm$; the tensor value is $0.311\,\fm$.
Table~\ref{tab:radiuscomparison} also gives the invariant $A$-slope radii.

\begin{table}[tbp]\centering\small
\begin{tabular}{lrrrrrr}\toprule
 &\multicolumn{2}{c}{$r_A$ [fm]}&\multicolumn{2}{c}{$r_{\rm mass,BF}$ [fm]}&\multicolumn{2}{c}{$D(0)$}\\\cmidrule(lr){2-3}\cmidrule(lr){4-5}\cmidrule(lr){6-7}Target & Full & HT & Full & HT & Full & HT\\\midrule
$0^{++}$ & 0.211 & 0.239 & 0.312 & 0.276 & -2.930 & -1.371\\
$0^{-+}$ & 0.264 & 0.264 & 0.296 & 0.300 & -2.408 & -2.685\\
$2^{++}$ (canonical spin) & 0.326 & 0.278 & 0.330 & 0.311 & -0.733 & -2.308\\\bottomrule
\end{tabular}
\caption{Full and HT forward observables at the same reference background. The tensor entries refer to the canonical-spin monopole. The HT physical trace equals the normalized scalar-source overlap. The
parameters are determined without GFF input. Pseudoscalar entries use
the reference infrared-decaying branch. Their propagated parameter errors
and separate branch sensitivity are given in Appendix~\ref{sec:calibration}.}
\label{tab:radiuscomparison}
\end{table}

The pseudoscalar $A$-slope radius, $r_{A,a}=0.264\,\fm$, is unchanged
between the full and HT responses. With the reference axion branch,
the HT radii have $0^{++}<0^{-+}<2^{++}$ in both listed conventions,
whereas the Full Breit radii place the pseudoscalar below the scalar.
The HT ordering is branch dependent: an infrared-growing axion solution
can give a pseudoscalar radius larger than the tensor radius.

\subsection{Predictions in the unfitted channels}
\label{sec:unfittedtests}
The pseudoscalar $A(Q^2)$ is identical in Full and HT. It tests the
geometry, axion kinetic function $Z(b)$, normalized target mode and
probe propagation. The two fitted masses leave a discrete branch
ambiguity even at $b_a=3$. The reference infrared-decaying branch gives
$r_{A,a}=0.264\,\fm$, $r_{{\rm mass,BF},a}=0.300\,\fm$ and
$D_{a,\HT}(0)=-2.685$. An infrared-growing solution at the same $b_a$,
$(\zeta_1,\zeta_2)=(13.5733,-32.3562)$, reproduces both masses and gives
$0.343\,\fm$, $0.369\,\fm$ and $-2.500$, respectively. Both branches
satisfy $Z>0$ and $\log Z=O(\log b)$.

A measurement of $A_a$ can therefore select the axion-sector response
before its trace and $D_a$ test the cubic coupling. On the growing branch
$D_{a,\rm Full}(0)=-2.486$ is close to HT, whereas the reference branch
gives $-2.408$ and $-2.685$. The tested branch ranges in
Appendix~\ref{sec:calibration} are distinct from the conditional parameter
errors. The tensor channel is unaffected by this axion ambiguity.

The tensor channel is more sensitive to the gravitational cubic vertices.
At the same background, Full and HT give
$r_{A,T}=0.326$ and $0.278\,\fm$, respectively, and
$\overline D_T(0)=-0.733$ and $-2.308$.
The $A$ difference is entirely the known prefactor in
Eq.~\eqref{eq:maintensorfull}: its squared-radius shift is exactly
$4(\hbar c)^2/M_T^2$, and at $Q^2=M_T^2$ the HT value is $30/11$
times the Full value. This provides a direct test of the finite-mass
coupling assignment on the same background. The lattice comparison uses
five diagonal equal-rest-spin matrix elements after rotationless boosts
to $p=(E_Q,0,0,-Q/2)$ and $p'=(E_Q,0,0,Q/2)$. Writing
$\overline{\mathcal T}^{\mu\nu}=\tfrac15\sum_{s=-2}^{2}
\langle p',s|T^{\mu\nu}|p,s\rangle$, the required combinations are
\[
 \overline A_T=\frac{\overline{\mathcal T}^{00}
                         +\overline{\mathcal T}^{11}}{2E_Q^2},
 \qquad
 \overline D_T=-\frac{2\overline{\mathcal T}^{11}}{Q^2}.
\]
Equation~\eqref{app:spinmonopolemap} also gives the trace projection.
Euclidean lattice matrix elements must be converted to this physical EMT
convention. The spin average is the canonical-spin monopole, rather than
a diagonal helicity average or one invariant spin-two GFF.

\subsection{Pressure and shear from the form factors}
\label{sec:stress}
We compute the Breit-frame pressure and shear by numerical Fourier
transformation of the holographic form factors. With
$E_Q=\sqrt{M_X^2+Q^2/4}$, the convention is
\begin{equation}
\begin{aligned}
 \mathcal D_X(r)&=\int\frac{\dd^3Q}{(2\pi)^3}
 e^{-i\mathbf Q\cdot\mathbf r}\frac{M_X}{E_Q}D_X(-Q^2),\\
 p_X(r)&=\frac{\mathcal D_X''+2\mathcal D_X'/r}{6M_X},\qquad
 s_X(r)=-\frac{\mathcal D_X''-\mathcal D_X'/r}{4M_X}.
\end{aligned}
\label{eq:mainstresses}
\end{equation}
The target-dependent recoil is retained in Full and HT. Conditionally
convergent Full integrals use the Abel prescription. Numerical transforms
subtract and restore the analytic ultraviolet tail; the forward stress
moments and conservation identities were checked without rescaling any
profile~\cite{polyakov_stress}. These checks include the von Laue condition
$4\pi\int_0^\infty\dd r\,r^2p_X(r)=0$.

Figure~\ref{fig:mainstressphysical} shows the weighted profiles $r^2p$ and
$r^2s$. Full scalar and tensor vertices generate $D\sim Q^{-2}$ and
positive leading $1/r^2$ pressure and shear cores, so their weighted
profiles approach nonzero limits. The Full tensor normal stress
$p+2s/3$ is negative over an inner interval, approximately
$0.03$--$0.25\,\fm$, although its asymptotic core is positive.
HT gives $D_S\sim Q^{-4}$ and $D_a,\overline D_T\sim Q^{-6}$.
Its scalar pressure is logarithmic, while the pseudoscalar and tensor
pressures are finite at the origin. All HT weighted stresses vanish there.

For $k=Q/m\to\infty$, $u_S\sim u_3\xi^3$ gives
$A_S\sim48u_3^2/k^4$ and $D_{S,\HT}\sim-16u_3^2/k^4$.
The corresponding central limits are
\begin{equation}
 \frac{p_{S,\HT}(r)}{m^4}
 =\frac{8u_3^2}{3\pi^2}\log\frac1{mr}+O(1)
 \simeq0.04760\log\frac1{mr}+O(1),\qquad
 \frac{s_{S,\HT}(0^+)}{m^4}=\frac{4u_3^2}{3\pi^2}\simeq0.02380.
 \label{eq:HTscalarcore}
\end{equation}
For the other two targets $s_{\HT}(0)=0$. These locally integrable central
structures characterize the model's all-transfer continuation in the
Breit convention. Their short-distance behavior probes momenta beyond
the available lattice range. Relativistic transverse densities require a
separately specified matrix element and normalization~\cite{jaffe2021,lorce2019,freese2022}.

\section{Conclusions}
\label{sec:conclusion}
We have derived the full constrained gravitational cubic vertices of a
BCE-inspired Einstein--scalar--axion reconstruction and calculated the
scalar, pseudoscalar and canonical-spin-averaged tensor glueball form
factors. Glueball spectra and Wilson observables determine the common
background and dimensional scale without GFF input. The Full and HT
calculations therefore test two cubic-coupling assignments on identical
radial dynamics.

The normalized HT weights are motivated by the leading fixed-probe,
large-target structure of the complete amplitudes. Their positive radial
measures imply positive, completely monotone $A$ and physical trace
form factors. The exact finite-mass EMT relation determines $D$, including
$-A/3$, and cancels isolated tensor-source poles from the trace.
For the 27 scalar lattice observations at the common calibrated point,
$\chi^2_{AD}/\mathrm{dof}=2.73$ for Full and $0.63$ for HT,
favoring the normalized HT couplings.
The HT scalar Breit mass radius is $0.276\,\fm$, consistent with
$0.263(31)\,\fm$ from the lattice GFF analysis.

The reference infrared-decaying axion branch gives a pseudoscalar HT
mass radius of $0.300\,\fm$ and $D_a(0)\simeq-2.69$; a growing branch
reproduces the same two masses with a larger radius. The shared
pseudoscalar $A$ can distinguish these axion-sector choices, after which
its trace and $D$ test the scalar-channel coupling. The tensor prediction,
$0.311\,\fm$ and $\overline D_T(0)\simeq-2.31$, more directly separates
the cubic prescriptions. The scalar and tensor forward results are stable
under the tested warp interpolation, and a two-percent relative-scale
variation preserves the scalar Full--HT contrast. The remaining
$r_0\sqrt{\sigma}$ discrepancy constrains the effective-string sector.

Numerical Fourier transforms give the exact-recoil Breit pressure and
shear distributions. HT softens the Full scalar and tensor stress cores;
the scalar retains a logarithmic pressure and finite central shear.
Lattice pseudoscalar and polarized tensor matrix elements can test these
couplings. Transition form factors and glueball scattering would extend
the comparison to contact and exchange interactions in the same
framework~\cite{bianchi2003three,muckprisco2004,muck2005}.

\acknowledgments
This work was supported by the U.S.~National Science Foundation under
Grant No.\ 2412625, by the U.S.~Department of Energy, Office of Science,
Office of Nuclear Physics under the umbrella of the Quark-Gluon
Tomography (QGT) Topical Collaboration under Contract DE-SC0023646, and
by DOE Grant No.\ DE-FG02-04ER41309.

\paragraph{Data Availability Statement.}
The model parameters and numerical form-factor, pressure and shear tables
will be deposited at Zenodo: https://doi.org/10.5281/zenodo.23188093.
The lattice inputs are available from the publications and data record
cited in the text.

\paragraph{Code Availability Statement.}
The routines for evaluating and plotting the reference-model form factors
and mechanical profiles will accompany the Zenodo deposit:
https://doi.org/10.5281/zenodo.23188093.

\FloatBarrier\clearpage\appendix\numberwithin{equation}{section}

\section{Radial action, source normalization, and boundary prescription}
\label{app:radial}

This appendix fixes the action, particle normalization and metric-source conventions. The full-vertex calculation starts from
\begin{equation}
 S=\Nfive\int\dd^5x\sqrt{-g}\left[R-\frac12K(b)(\partial b)^2
 -\frac12Z(b)(\partial a)^2-V(b)\right]+S_{\rm GH}+S_{\rm ct}.
 \label{app:action}
\end{equation}
The bulk signature is $(-,+,+,+,+)$. The conformal background is
$\dd s_E^2=e^{2A_E(z)}(\dd z^2+\eta^{\rm bulk}_{\mu\nu}\dd x^\mu\dd x^\nu)$,
with $\bar a=0$. For radial formulas introduce $\xi=mz$,
$A(\xi)=A_E(z)-\log(Lm)$, $\mathcal V=L^2V$, and set $L=m=1$
after rescaling the coordinates. In this and the next appendix, an unqualified radial prime means
$\dd/\dd\xi$; the proper-radial ADM derivation explicitly uses dots. Define
\begin{equation}
 \phi(b)=\int_0^b\sqrt{K(t)}\,\dd t,\qquad
 P=e^{3A},\qquad c=-\frac{\bar\phi'}{A'},\qquad
 x=\frac{Q^2}{m^2},\qquad\mu_X=\frac{M_X^2}{m^2}.
 \label{app:variables}
\end{equation}
Thus $\mu_X$ here is dimensionless, whereas a dimensional formula using
$M_X^2$ must retain the factors of $m$. The radial normalization absorbs
$\sqrt{\Nfive}$ into each canonical four-dimensional particle field;
this fixes the normalization before the overlaps are evaluated.

\subsection{Background equations and the two scalar charts}
The dynamical model fixes its coordinate field by $\bar b=\xi$ and writes
$A=-\log\xi+F(\bar b)$. Einstein's equations then reconstruct
\begin{align}
 K(b)&=6\left[F_{,b}^2-F_{,bb}-\frac{2F_{,b}}b\right],\notag\\
 \mathcal V(b)&=-3e^{-2F}
 \left[4-6bF_{,b}+b^2F_{,bb}+3b^2F_{,b}^2\right].
 \label{app:reconstruction}
\end{align}
They imply $\bar\phi'^2=6(A'^2-A'')$ and
$e^{2A}\mathcal V=-3(A''+3A'^2)$; the scalar equation follows on a
monotone trajectory. All potential derivatives in the cubic vertices are
\emph{canonical} derivatives:
\begin{align}
 \mathcal V_{,\phi\phi}
 &=\frac{\mathcal V_{,bb}}K-\frac{K_{,b}\mathcal V_{,b}}{2K^2},\notag\\
 \mathcal V_{,\phi\phi\phi}
 &=\frac{\mathcal V_{,bbb}}{K^{3/2}}
 -\frac{3K_{,b}\mathcal V_{,bb}}{2K^{5/2}}
 +\left[\frac{K_{,b}^2}{K^{7/2}}
 -\frac{K_{,bb}}{2K^{5/2}}\right]\mathcal V_{,b},\qquad
 Z_{,\phi}=\frac{Z_{,b}}{\sqrt K}.
 \label{app:canonicalderivatives}
\end{align}
For $F=-b^2/36+f_4b^4+\cdots$, $K(0)=1$ and
$\mathcal V(\phi)=-12-3\phi^2/2-\phi^4/12+O(\phi^6)$.

\subsection{Quadratic actions and source columns}
Let $q$ be the canonical active-scalar fluctuation in flat gauge, or its
gauge-invariant continuation, and $E_{\mu\nu}$ the TT metric fluctuation.
With $v=q/c$, the three quadratic actions are
\begin{align}
 S_q^{(2)}&=-\frac{\Nfive}{2}\int\dd\xi\dd^4x\,Pc^2
       [v'^2+\partial_\mu v\partial^\mu v],\notag\\
 S_a^{(2)}&=-\frac{\Nfive}{2}\int\dd\xi\dd^4x\,PZ
       [a'^2+\partial_\mu a\partial^\mu a],\notag\\
 S_E^{(2)}&=-\frac{\Nfive}{4}\int\dd\xi\dd^4x\,P
       \operatorname{tr}[E'^2+\partial_\mu E\partial^\mu E].
 \label{app:kinetic}
\end{align}
The first expression includes the usual choice of reduction surface.
Equivalently its bulk integrand is
$P[q'^2+(\partial q)^2+\mathcal U_q q^2]$, where
$\mathcal U_q=c''/c+3A'c'/c$, with the explicit added surface
$\Nfive[Pc'q^2/c]_{\rm UV}^{\rm IR}/2$.
Writing $E_{\mu\nu}=\sqrt2\,\epsilon_{\mu\nu}t$ with
$\epsilon_{\mu\nu}\epsilon^{\mu\nu}=1$ gives a canonical scalar
normalization for each tensor polarization. The mode equations are
\begin{equation}
 -(w_X f_{X,n}')'=\mu_{X,n}w_Xf_{X,n},\qquad
 (f_X,w_X)=(q/c,Pc^2),(a,PZ),(t,P).
 \label{app:SL}
\end{equation}
The ground-state radial functions used in matrix elements obey
\begin{equation}
 \int\dd\xi\,P u_S^2=1,\qquad
 \int\dd\xi\,PZ u_a^2=1,\qquad
 \int\dd\xi\,P u_T^2=2.
 \label{app:norms}
\end{equation}
Their ultraviolet powers are $u_S\sim u_3\xi^3$ and
$u_a\sim u_{4,a}\xi^4$, $u_T\sim C_T\xi^4$.
Here $u_S$ is the coupled scalar--metric mode whose tower is sourced by
$J_q=j_\phi+m j_\sigma$, not an independent probe-scalar target; the
functional trace insertion is derived in Appendix~\ref{sec:sourcemap}.
The source columns satisfy
\begin{equation}
 (PH_T')'=xPH_T,\qquad(Pc^2H_S')'=xPc^2H_S,
 \qquad H_T(0,x)=H_S(0,x)=1.
 \label{app:sources}
\end{equation}
On the dynamical half-line choose the regular source and square-integrable
mode branches; a numerical upper endpoint is not a hard wall. The
Schr\"odinger potentials are $U_X=(\sqrt{w_X})''/\sqrt{w_X}$.
For $A\sim-y_{\rm IR}\xi^2/2$, $c^2\to6$ and
$\log Z=O(\log\xi)$, their half-line oscillator limit gives
$\mu_{X,n}/n\to6y_{\rm IR}$.

\subsection{Metric and scalar boundary variations}
\label{app:boundaryfunctional}
The regulated Dirichlet functional includes
\begin{equation}
 \begin{gathered}
 S_\partial=2\Nfive\int_{\partial\mathcal M}\dd^4x\sqrt{-\gamma}\,K
          +S_{\rm ct}+S_{\rm fin}+S_{\rm IR},\\
 \pi_\phi=-\Nfive\sqrt{-\gamma}\,n^M\partial_M\phi,\qquad
 \pi_a=-\Nfive\sqrt{-\gamma}\,Z n^M\partial_M a.
 \end{gathered}
 \label{app:boundarymomenta}
\end{equation}
The extrinsic-curvature orientation is fixed by the spacelike-normal
ADM convention in Appendix~\ref{sec:ESAADM}. Counterterm derivatives
are added to these momenta before the renormalized source responses are
formed. $S_{\rm fin}$ fixes the finite local source convention, and
$S_{\rm IR}$ specifies the infrared domain; the dynamical half-line
uses the regular variational domain without an additional infrared interaction.
The independent source variation is
\begin{equation}
 \delta W_{\rm ren}=\int\dd^4x\sqrt{-g_{(0)}}
 \left[\frac12T_+^{\mu\nu}\delta g_{(0)\mu\nu}
       +\mathcal O_\phi\delta j_\phi+\mathcal O_a\delta j_a\right].
 \label{app:boundaryvariation}
\end{equation}
The background mass source is held fixed when the additional sources
are subsequently set to zero. Local subtraction follows the canonical
potential and kinetic functions of the specified action
\cite{Bianchi:2001ren,Skenderis:2002ren}.

The local counterterms follow the reconstructed canonical action.
Their source-only contributions have no double-target-pole residue for
the normalizable modes considered here. The scalar-source chain rule
and the radial reduction surfaces are retained before pole extraction.

\subsection{Boundary and forward prescriptions}
The Gibbons--Hawking term and local ultraviolet source counterterms in
Eq.~\eqref{app:action} define the variational problem before particle
poles are extracted. With two normalizable external legs their local
contact terms and the scalar reduction surface have vanishing residues
in the matrix elements used here. The auxiliary surface-separated tensor representation differs by the
radial derivatives
\begin{equation}
 \tfrac14\partial_\xi[P\bar\phi' q\operatorname{tr}(E^2)],
 \qquad\partial_\xi[PA'\operatorname{tr}(E^3)].
 \label{app:TTsurfaces}
\end{equation}
Their full double-pole residues vanish on the normalizable domains.
The full response includes these terms before the normalizable domain is imposed.
The HT construction uses the resulting on-shell amplitude and the target-mode
identity in Appendix~\ref{app:HTderivation}.
The axion flux surfaces vanish on the dynamical half-line. The boundary
functional is retained before extracting the particle residues.

For a nonsingular forward evaluation set $H_Y=1+xh_Y+O(x^2)$, where
\begin{equation}
 h_Y'(\xi)=-\frac{1}{w_Y(\xi)}\int_\xi^{\xi_{\rm IR}}w_Y(t)\dd t,
 \qquad h_Y(0)=0,\qquad (w_T,w_S)=(P,Pc^2).
 \label{app:forwardprobe}
\end{equation}
Here $\xi_{\rm IR}=\infty$.
The vanishing infrared flux selects this solution. Differentiated source
overlaps, rather than a difference of nearly equal amplitudes, determine
$D(0)$ below.

\subsection{Conformal constraints and the canonical scalar fluctuation}
\label{app:constraints}
The local vertices are most simply written in the canonical chart
$\phi=\phi(b)$. Let $n$ be the fractional radial lapse perturbation and
$\chi$ the conformal shift potential. With all four-dimensional
contractions taken in $\eta^{\rm bulk}$, define
\begin{equation}
 a_r=\frac{\bar\phi'}{6A'}=-\frac c6,\qquad
 C_{\mu\nu}=\partial_\mu\partial_\nu\chi,\qquad C=\Box\chi.
 \label{app:constraintnotation}
\end{equation}
The first-order Hamiltonian and momentum constraints
in flat-slice gauge give
\begin{equation}
 n=a_rq,\qquad
 \Box\chi=a_rq'-a_r'q,\qquad
 \chi(\xi,k)=\frac{a_r'q-a_rq'}{k^2},\qquad \Box=-k^2.
 \label{app:linearconstraints}
\end{equation}
Thus the lapse and shift are not extra propagating target modes.
A scalar source column $q_s=cH_S$ has
$n_s=-c^2H_S/6$ and $x\chi_s=c^2H_S'/6$. On a scalar target pole
$k^2=-\mu_S$, their radial columns reduce to $N=-cu/6$ and
$B=-(cu'-c'u)/(6\mu_S)$, respectively. The apparent inverse-momentum
singularities are evaluated with the differentiated source prescription
\eqref{app:forwardprobe}.

The canonical bulk mass in the $q$ equation can equivalently be written
\begin{equation}
 \mathcal U_q=e^{2A}\left[
 \mathcal V_{,\phi\phi}
 +\frac{2\bar\phi'\mathcal V_{,\phi}}{3A'}
 +\frac{\mathcal V\bar\phi'^2}{9A'^2}\right]
 =\frac{c''}{c}+3A'\frac{c'}c .
 \label{app:activemass}
\end{equation}
The equality follows from the background equations and includes the
scalar--metric mixing. In particular, a probe equation containing only
$\mathcal V_{,\phi\phi}$ would be a different response problem. These
relations are the conformal-coordinate identities of the retained
Einstein--scalar--axion system, evaluated here on the reconstructed
$\mathcal V(\phi),Z(\phi)$. Appendix~\ref{sec:ESAADM} derives the constraints.

\subsection{From the active scalar to the metric trace source}
\label{sec:sourcemap}
This subsection derives the source change underlying the trace kernels. It fixes which
functional differentiation produces the trace insertion. The argument
uses the single active scalar with $\bar a=0$, the chosen standard
source branch, and the common variational boundary prescription.

\subsubsection{Gauge-invariant source combination}
Write the scalar slice-metric perturbation as
$2\psi\eta^{(+)}_{\mu\nu}$, alongside its TT and longitudinal parts.
In proper-radial notation $H=\dot{\mathcal A}$,
$v=\dot{\bar\phi}$ and $c=-v/H=-\bar\phi'/A'$, where dots in this
subsection denote proper-radial derivatives. The active invariant is
\begin{equation}
 q=\delta\phi-\frac vH\psi=\delta\phi+c\psi.
 \label{eq:mapgaugeinvariant}
\end{equation}
Indeed the radial-gauge shifts of $\delta\phi$ and $\psi$ are
proportional to $v$ and $H$, so their combination cancels. In flat-slice
gauge $\psi=0$, the canonical fluctuating scalar equals $q$; the Weyl
source is then encoded in its boundary coefficient, not set to zero.
Let $j_\sigma$ be the boundary Weyl source,
$\delta g_{(0)\mu\nu}=2j_\sigma\eta^{(+)}_{\mu\nu}$, and $j_\phi$
the independent additional mass source in the normalization
$\delta\phi=zj_\phi+\cdots$. Since $c=mz+\cdots$,
\begin{equation}
 J_q\equiv\lim_{z\to0}\frac qz=j_\phi+m j_\sigma.
 \label{eq:mapsourcecombination}
\end{equation}
The independent sources are differentiated before they are set to their
background values. A pure trace insertion holds $j_\phi=0$ fixed;
it does not replace a metric variation by an arbitrary scalar probe.

\subsubsection{The source profile and the constrained scalar operator}
Using the dimensionless radial coordinate $\xi=mz$ for derivatives,
the flat-gauge scalar equation contains
$\mathcal U_q=c''/c+3A'c'/c$. Substitution of $q=c\zeta$ into that
equation gives
\begin{equation}
 (Pc^2\zeta')'=xPc^2\zeta,
 \qquad
 q(z,Q)=\frac{c(z)}m H_S(z,Q)J_q,
 \qquad H_S(0,Q)=1.
 \label{eq:mapprofile}
\end{equation}
Consequently the dimension-one scalar source kernel is
$\mathcal K_\phi=cH_S/m$, with
$\mathcal K_\phi/z\to1$. The column generated by a unit Weyl source is
\begin{equation}
 q_s=cH_S,\qquad n_s=-\frac{c^2}{6}H_S,
 \qquad x\chi_s=\frac{c^2}{6}H'_S .
 \label{eq:maptracecolumns}
\end{equation}
The last two expressions follow from the ADM constraints in
Appendix~\ref{sec:ESAADM} and their conformal form
in Eq.~\eqref{app:linearconstraints}. The radial inverse-transfer
combination has the finite limit
\begin{equation}
 \left.\frac{H_S'}x\right|_{x=0}=h_S',\qquad
 h_S'(\xi)=-\frac1{Pc^2}\int_\xi^{\xi_{\rm IR}}Pc^2\,\dd v.
 \label{eq:mapforwardquotient}
\end{equation}
The unit forward profile is $H_S(\xi,0)=1$, but the quotient in
Eq.~\eqref{eq:mapforwardquotient} is not zero. All profiles use the
same infrared domain as their quadratic operator.

\subsubsection{Functional differentiation, trace sign and particle poles}
In the mostly-plus convention the renormalized source variation is
\begin{equation}
 \delta W=\int\dd^4x\sqrt{-g_{(0)}}
 \left[\tfrac12 T_+^{\mu\nu}\delta g_{(0)\mu\nu}
       +\mathcal O_\phi\delta j_\phi+\mathcal O_a\delta j_a\right].
 \label{eq:mapsourcevariation}
\end{equation}
A variation of $j_\sigma$ therefore inserts the contracted metric response.
Writing $\widetilde W[J_q,j_\sigma,\ldots]=W[j_\phi=J_q-m j_\sigma,j_\sigma,\ldots]$,
where the other independent sources are unchanged, gives the chain rule
\begin{equation}
 \begin{aligned}
 \left.\frac{\delta W}{\delta j_\sigma}\right|_{j_\phi}
 &=m\frac{\delta\widetilde W}{\delta J_q}
  +\left.\frac{\delta\widetilde W}{\delta j_\sigma}\right|_{J_q},\\
 \left.\frac{\delta W}{\delta j_\phi}\right|_{j_\sigma}
 &=\frac{\delta\widetilde W}{\delta J_q}.
 \end{aligned}
 \label{eq:mapchainrule}
\end{equation}
The explicit terms include the transformed source and boundary
functional and must be retained in the unamputated source problem.
For the separated target matrix elements, source-only local terms have
no double target pole; the reduction surfaces must still be evaluated
on the stated normalizable domain. This is the condition under which
the source columns in Eq.~\eqref{eq:maptracecolumns} implement the
trace calculation. It is not a general identity equating the full
renormalized Yang--Mills trace to a bare dimension-three operator.

The displayed physical amplitudes use $\eta^{\rm phys}=-\eta^{(+)}$.
With the common source normalization, the physical trace is the negative
of the contracted mostly-plus response, as in Eq.~\eqref{app:traceD}.
For a source amplitude bilinear in two target modes, one inserts the
column $(q_s,n_s,\chi_s)$ into the appropriate cubic density, sums all
identical-field assignments, and extracts the coefficient of the two
target poles. Canonical particle normalization divides out their residues.
The resulting scalar trace is
\begin{equation}
 \widehat\Theta_{S,\rm full}=-\int\dd\xi\,P
 [(F_0+xF_1+x^2F_2)H_S+(G_0+xG_1)H_S'],
 \qquad A_S^S=\widehat\Theta_S/(2\mu_S).
 \label{eq:maptraceamplitude}
\end{equation}
The axion and tensor targets use the same trace-source column and their
own normalized external modes. A separate axion source $j_a$ is not
needed for an axion-target GFF; its topological normalization is a
separate source-matching datum.

At zero transfer the full source problem has $A_X^{\rm S}(0)=1$.
The HT model uses the same probe functions and identifies its normalized
scalar overlap directly with the physical trace,
$A_X^{\rm S}=\mathcal S_X$, through Eq.~\eqref{eq:HTtracecompletion}. Appendix~\ref{app:HTderivation} gives the on-shell
large-target extraction and normalization step.

\section{Full constrained cubic vertices}
\label{app:kernels}

Use $\eta^{\rm phys}=(+,-,-,-)=-\eta^{\rm bulk}$ for the physical EMT.
For spin zero, $\Delta=p'-p$, $\bar p=(p+p')/2$, and $t=-Q^2$,
\begin{equation}
 \langle p'|T^{\mu\nu}|p\rangle=2\bar p^\mu\bar p^\nu A(t)
 +\tfrac12(\Delta^\mu\Delta^\nu-\eta_{\rm phys}^{\mu\nu}t)D(t).
 \label{app:EMT}
\end{equation}
Denote the physical trace divided by $m^2$ by $\widehat\Theta$.
It is minus the contracted mostly-plus bulk response. The convention is
\begin{align}
 A^{\rm S}(x)&=\frac{\widehat\Theta(x)}{2\mu},\qquad
 D(x)=\frac{2\widehat\Theta(x)-(4\mu+x)A(x)}{3x},\notag\\
 A(0)&=A^{\rm S}(0)=1,\qquad
 D(0)=\frac{2\dot{\widehat\Theta}(0)-1-4\mu\dot A(0)}3.
 \label{app:traceD}
\end{align}
Dots mean $\dd/\dd x$. The same last definitions apply to the specified
canonical-spin tensor average, with bars on its form factors.

\subsection{Deriving the constrained action in proper radial coordinates}
This derivation applies to the canonical action with arbitrary
$V(\phi),Z(\phi)$ and a zero-axion background. Here $r$ is proper radial
distance, $\mathcal A(r)=A_E(z(r))$, and an overdot denotes $\dd/\dd r$.
Write $H=\dot{\mathcal A}$ and $v=\dot{\bar\phi}$. The background equations are
\begin{equation}
 \dot H=-\frac{v^2}{6},\qquad
 \dot v+4Hv=V_{,\phi},\qquad
 H^2=\frac{v^2}{24}-\frac V{12}.
 \label{eq:ESAbgEquations}
\end{equation}
The potential is dimensional in these proper-radial formulas; the
rescaled conformal densities below use $\mathcal V=L^2V$.
The reconstructed background obeys this system with its canonical potential.
\subsubsection{Radial ADM formulation and the zero-axion fluctuation problem}
\label{sec:ESAADM}
Use a spacelike radial normal and
\begin{equation}
 \begin{gathered}
 \dd s^2=N^2\dd r^2+
 \gamma_{\mu\nu}(\dd x^\mu+N^\mu\dd r)(\dd x^\nu+N^\nu\dd r),\\
 K_{\mu\nu}=\frac{\dot\gamma_{\mu\nu}-\nabla_\mu N_\nu-\nabla_\nu N_\mu}{2N},
 \qquad D_r=\partial_r-N^\mu\partial_\mu .
 \end{gathered}
 \label{eq:ESAADMmetric}
\end{equation}
After the gravitational boundary term is included, the exact bulk action is

\begin{equation}
\begin{split}
 \frac{\mathcal L_{\rm ADM}}{\mathcal N_5}=N\sqrt{-\gamma}\Big[&
 R[\gamma]+K^2-K_{\mu\nu}K^{\mu\nu}
 -\frac{(D_r\phi)^2+Z(\phi)(D_ra)^2}{2N^2}\\
 &-\frac12\gamma^{\mu\nu}
 (\partial_\mu \phi\partial_\nu \phi+Z\partial_\mu a\partial_\nu a)-V\Big].
 \end{split}
 \label{eq:ESAADMaction}
\end{equation}

The extrinsic-curvature sign follows from the spacelike normal and gives the constraint in Eq.~\eqref{eq:ESAbgEquations}. This nonlinear ADM action also determines the remaining pure-gravity and mixed metric vertices; the scalar-insertion vertices derived below do not replace it.

Specialize the background to $\bar a=0$. Before gauge fixing, the slice-metric fluctuation contains the Weyl component $2\psi\eta^{(+)}_{\mu\nu}$, a transverse-traceless component $e_{\mu\nu}^{TT}$, and longitudinal components, all multiplied by $e^{2\mathcal A}$. The active gauge-invariant variable and the flat-slice gauge are
\begin{equation}
\begin{gathered}
 \mathfrak q=\delta \phi-\frac vH\psi,\qquad \psi=0,\qquad q=\mathfrak q,\\
 N=1+n,\qquad N^\mu=\partial^\mu\beta.
 \end{gathered}
 \label{eq:ESAgauge}
\end{equation}
The axion fluctuation is gauge invariant at this order because its background is constant. Variation with respect to the lapse and shift gives
\begin{equation}
 \begin{aligned}
 n&=a_rq,\qquad a_r=\frac v{6H},\\
 B\equiv\Box\beta&=\frac{2Vn-v\dot q+V_{,\phi}q}{6H}
                =-a_r\dot q+\dot a_rq .
 \end{aligned}
 \label{eq:ESAconstraints}
\end{equation}
A TT field or axion fluctuation alone does not source these first-order auxiliary fields. $H\ne0$ is assumed in this coordinate patch. The inverse $\Box$ uses the chosen common source prescription; zero transfer is defined by a smooth limit or the unsolved constraint.

Up to the boundary term retained in the variational functional, the active quadratic action is
\begin{equation}
 \begin{gathered}
 S_q^{(2)}=-\frac{\mathcal N_5}{2}\int\dd r\dd^4x\,e^{4\mathcal A}
 [\dot q^2+e^{-2\mathcal A}(\partial q)^2+\mathcal U_{\phi}q^2],\\
 \mathcal U_\phi=V_{,\phi\phi}+\frac{2vV_{,\phi}}{3H}+\frac{Vv^2}{9H^2}.
 \end{gathered}
 \label{eq:ESAactiveQuadratic}
\end{equation}
The last two terms in $\mathcal U_\phi$ are fixed by the metric constraints. The axion and tensor actions are
\begin{equation}
 \begin{aligned}
 S_a^{(2)}&=-\frac{\mathcal N_5}{2}\int\dd r\dd^4x\,e^{4\mathcal A}Z
 [\dot a^2+e^{-2\mathcal A}(\partial a)^2],\\
 S_{TT}^{(2)}&=-\frac{\mathcal N_5}{4}\int\dd r\dd^4x\,e^{4\mathcal A}
 [\dot e_{\mu\nu}\dot e^{\mu\nu}
 +e^{-2\mathcal A}\partial_\rho e_{\mu\nu}\partial^\rho e^{\mu\nu}].
 \end{aligned}
 \label{eq:ESAinertQuadratic}
\end{equation}
Indices and $\Box$ in these fluctuation formulas use $\eta^{(+)}$. Only first-order lapse and shift solutions are needed in the cubic bulk action: their second-order corrections multiply the solved first-order constraints. Surface contributions are kept separately.

\subsubsection{A direct derivation of the active tensor vertex}
\label{sec:ESAtensorDerivation}
Hold the contravariant shift $N^\mu=\partial^\mu\beta$ fixed while varying
$\gamma_{\mu\nu}=e^{2\mathcal A}(\eta^{(+)}_{\mu\nu}+e_{\mu\nu})$.
Set $B_{\mu\nu}=\partial_\mu\partial_\nu\beta$ and let the matrix $\mathcal E^\mu{}_\nu=NK^\mu{}_\nu$. To first order in the transverse-traceless field,

\begin{equation}
\begin{gathered}
 \mathcal E=H\mathbf1-B+\frac12\dot e
 -\frac12\partial^\rho\beta\,\partial_\rho e
 -\frac12[B,e]+O(e^2),\\
 \operatorname{tr}\mathcal E=4H-B.
 \end{gathered}
 \label{eq:ESAextrinsicExpansion}
\end{equation}

The linear TT variation of the curvature $R[\gamma]$ and of $\sqrt{-\gamma}$ vanishes. The extrinsic term is
$[(\operatorname{tr}\mathcal E)^2-\operatorname{tr}(\mathcal E^2)]/N$.
Its contribution linear in $e$ is
\begin{equation}
 \frac1N\left[B^{\mu\nu}\dot e_{\mu\nu}
     -B^{\mu\nu}\partial^\rho\beta\,\partial_\rho e_{\mu\nu}\right].
 \label{eq:ESAextrinsicTT}
\end{equation}
The term without $n$ in $B^{\mu\nu}\dot e_{\mu\nu}$ is a boundary-coordinate total derivative by TT transversality. The commutator in Eq.~\eqref{eq:ESAextrinsicExpansion} drops from the trace contractions. Expanding $1/N=1-n+\cdots$ and varying the scalar slice kinetic term leaves
\begin{equation}
 \boxed{\begin{aligned}
 \frac{\mathcal L_{eqq}^{(3)}}{\mathcal N_5e^{4\mathcal A}}
 ={}&\frac12e^{-2\mathcal A}e^{\mu\nu}\partial_\mu q\partial_\nu q
 -nB^{\mu\nu}\dot e_{\mu\nu}\\
 &-B^{\mu\nu}\partial^\rho\beta\,\partial_\rho e_{\mu\nu}.
 \end{aligned}}
 \label{eq:ESAeqq}
\end{equation}
This expression is obtained directly from the closed action at the displayed order in fluctuations. It is first order in the tensor field and second order in active scalar--metric fluctuations. The last two terms are gravitational contributions; no tensor potential term has been omitted. No integration by parts in the radial coordinate is used to reach Eq.~\eqref{eq:ESAeqq}.

\subsubsection{The active cubic and the axion interactions}
\label{sec:ESAcubics}
The active self-interaction follows independently from the same ADM action. At $e=a=0$, define $X_q=\partial_\mu\beta\partial^\mu q$. The flat-gauge expression before expansion is

\begin{equation}
\begin{split}
 \frac{\mathcal L_{\rm flat}}{\mathcal N_5e^{4\mathcal A}}={}&
 \frac{12H^2-6HB+B^2-B_{\mu\nu}B^{\mu\nu}}{1+n}
 -\frac{(v+\dot q-X_q)^2}{2(1+n)}\\
 &-(1+n)\left[\frac12e^{-2\mathcal A}(\partial q)^2+V(\bar \phi+q)\right].
 \end{split}
 \label{eq:ESAflatAction}
\end{equation}

Use $12H^2-v^2/2=-V$, expand through cubic degree, and substitute Eq.~\eqref{eq:ESAconstraints}. The result is

\begin{equation}
 \begin{split}
 \frac{\mathcal L_{qqq}^{(3)}}{\mathcal N_5e^{4\mathcal A}}={}&
 Vn^3-6Hn^2B-n(B^2-B_{\mu\nu}B^{\mu\nu})-vn^2\dot q
 +\frac n2[\dot q^2-e^{-2\mathcal A}(\partial q)^2]\\
 &+(\dot q-vn)\partial_\mu\beta\partial^\mu q
 -\frac n2V_{,\phi\phi}q^2-\frac16V_{,\phi\phi\phi}q^3.
 \end{split}
 \label{eq:ESAqqq}
\end{equation}

This active self-interaction supplies the scalar-target trace kernel. All coefficients remain exact functions of $\bar \phi(r)$.

Expanding the axion part of Eq.~\eqref{eq:ESAADMaction} gives
\begin{equation}
 \begin{split}
 \frac{\mathcal L_{qaa}^{(3)}}{\mathcal N_5e^{4\mathcal A}}={}&
 -\frac12qZ_{,\phi}[\dot a^2+e^{-2\mathcal A}(\partial a)^2]\\
 &+\frac12nZ[\dot a^2-e^{-2\mathcal A}(\partial a)^2]
 +Z\dot a\partial_\mu\beta\partial^\mu a,
 \end{split}
 \label{eq:ESAqaa}
\end{equation}
while a tensor probe has
\begin{equation}
 \frac{\mathcal L_{eaa}^{(3)}}{\mathcal N_5e^{4\mathcal A}}
 =\frac12e^{-2\mathcal A}Z(\bar \phi)e^{\mu\nu}\partial_\mu a\partial_\nu a.
 \label{eq:ESAeaa}
\end{equation}
For the reconstructed $Z$ the direct $qaa$ coupling is
$-Z_{,\phi}q(\partial a)^2/2$. Its lapse and shift partners follow from the same action. At $\bar a=0$ the $a\mapsto-a$ symmetry forbids vertices with an odd number of axion legs, including $aaa$ and $qqa$. Equations~\eqref{eq:ESAeqq}--\eqref{eq:ESAeaa} are the complete scalar-target bulk vertices for the insertions considered here. Purely gravitational terms and terms with two external metric legs are determined by the exact ADM action and are not replaced by this selected list.

\paragraph{Conversion to the conformal convention.}\label{app:ADMconversion}
For $\dd r=-e^{A_E}\dd z$ one has
$H=-e^{-A_E}\partial_z A_E$, $v=-e^{-A_E}\partial_z\bar\phi$ and
$\beta=e^{-A_E}\chi$. After rescaling to $\xi=mz$, the proper-radial constraint
$\Box\beta=-a_r\dot q+\dot a_rq$ becomes
$\Box\chi=a_rq'-a_r'q$. In particular the proper-radial term
$-nB^{\mu\nu}\dot e_{\mu\nu}$ becomes the positive conformal term
$nC^{\mu\nu}E'_{\mu\nu}$, while the scalar shift-advection term becomes
$-(q'-\bar\phi'n)\partial_\mu\chi\partial^\mu q$.
The conformal shift in the line element is $N_z^\mu=-\partial^\mu\chi$. This change of coordinate orientation explains the signs in the
following common-measure densities.

\subsection{Constrained cubic action before elastic contraction}
\label{app:cubicdensities}
In the canonical chart, the bulk cubic action has the common measure
\begin{equation}
 S^{(3)}_{\rm bulk}=\Nfive\int\dd\xi\dd^4x\,P
 \bigl(\ell_{qqq}+\ell_{Eqq}+\ell_{qaa}+\ell_{Eaa}
                   +\ell_{qEE}+\ell_{EEE}\bigr).
 \label{app:sixvertices}
\end{equation}
For the active scalar and a tensor probe, the unsimplified densities are
\begin{align}
 \ell_{qqq}={}&e^{2A}\mathcal V n^3+6A'n^2C
 -n(C^2-C_{\mu\nu}C^{\mu\nu})-\bar\phi'n^2q'
 +\frac n2\bigl[q'^2-(\partial q)^2\bigr]\notag\\
 &-(q'-\bar\phi'n)\partial_\mu\chi\partial^\mu q
 -\frac{e^{2A}}2 n\mathcal V_{,\phi\phi}q^2
 -\frac{e^{2A}}6\mathcal V_{,\phi\phi\phi}q^3,
 \label{app:qqqvertex}\\
 \ell_{Eqq}={}&\frac12 E^{\mu\nu}\partial_\mu q\partial_\nu q
 +n C^{\mu\nu}E'_{\mu\nu}
 -C^{\mu\nu}\partial^\rho\chi\partial_\rho E_{\mu\nu}.
 \label{app:Eqqvertex}
\end{align}
The constraints \eqref{app:linearconstraints} are inserted after this
expansion. These are the conformal versions of the canonical ADM
vertices derived in Appendix~\ref{sec:ESAADM}, with canonical-potential
derivatives evaluated as in Eq.~\eqref{app:canonicalderivatives}. The vertices are evaluated from the reconstructed action and its canonical couplings. The lapse term in \eqref{app:Eqqvertex} is positive, whereas
the shift-advection term in \eqref{app:qqqvertex} is negative in this
conformal convention.

The other four densities are displayed below. They exhaust the cubic
bulk channels among $E,q,a$ at $\bar a=0$: terms with an odd number of
axion fields vanish by $a\mapsto-a$. Required boundary terms accompany Eq.~\eqref{app:sixvertices}.
For the two- and three-tensor channels the unintegrated ADM expressions
in Eq.~\eqref{eq:tensorunreduced} complete these densities.
We insert one source and two target columns, sum identical-field
assignments, and resolve regular momentum coefficients before evaluating the full response.

\subsection{Active scalar}
In this subsection set $u=u_S$, $\mu=\mu_S$,
$d=cu'-c'u$, $N=-cu/6$, $B=-d/(6\mu)$, and
$E_j=e^{2A}\mathcal V_{,\phi^j}$. The exact constrained cubic kernels give
\begin{align}
 A_{S,\rm full}&=\int\dd\xi\,P[H_Tu^2-2H_T'NB+xH_TB^2],\notag\\
 \widehat\Theta_{S,\rm full}&=-\int\dd\xi\,P
 [(F_0+xF_1+x^2F_2)H_S+(G_0+xG_1)H_S'],
 \label{app:scalarfull}
\end{align}
where the complete elastic polynomials, including the potential and
constraint contributions, are
\begin{align}
 F_0={}&-\frac1{36}\big[-2A'c^4uu'-4A'c^3c'u^2+E_0c^4u^2
 -18E_2c^2u^2+36E_3cu^2\notag\\[-2pt]
 &\hspace{4em}+6\mu c^2u^2+6c^2u'^2+24cc'uu'-12c'^2u^2\big],\notag\\
 F_1={}&\frac{9\mu c^2u^2+(18-c^2)d^2}{108\mu},\qquad
 F_2=-\frac{c^2d^2}{432\mu^2},\notag\\
 G_0={}&\frac{cu}{108}
 [3A'c^3u+2c^3u'-2c^2c'u-90cu'+36c'u],\qquad
 G_1=\frac{cu(c^2-36)d}{216\mu}.
 \label{app:scalarpolynomials}
\end{align}
The polynomials are the complete six-assignment contraction of the
canonical action. The normalized HT model is derived and defined
separately in Appendix~\ref{app:HTderivation}.

\subsection{Pseudoscalar}
For this target the common-measure cubic densities are
\begin{align}
 \ell_{qaa}&=-\tfrac12qZ_{,\phi}[a'^2+(\partial a)^2]
 +\tfrac12nZ[a'^2-(\partial a)^2]-Za'\partial_\mu\chi\partial^\mu a,
 \notag\\
 \ell_{Eaa}&=\tfrac12ZE^{\mu\nu}\partial_\mu a\partial_\nu a.
 \label{app:axionvertices}
\end{align}
Their action is $\Nfive\int P\ell\,\dd\xi\dd^4x$. On a unit trace
source $q_s=cH_S$, $n_s=-c^2H_S/6$, and $x\chi_s=c^2H_S'/6$.
Adding the two target assignments gives
\begin{align}
 \alpha_\pm&=-c^2Z/6\ \mathord{\pm}\ cZ_{,\phi},\notag\\
 \mathcal E_0&=\alpha_-u_a'^2+\mu_a\alpha_+u_a^2,\qquad
 \mathcal E_1=\alpha_+u_a^2/2,\qquad
 \mathcal J_0=-c^2Zu_au_a'/6,\notag\\
 A_{a,\rm full}&=\int\dd\xi\,PZH_Tu_a^2,\notag\\
 \widehat\Theta_{a,\rm full}&=-\int\dd\xi\,P
 [ (\mathcal E_0+x\mathcal E_1)H_S+\mathcal J_0H_S'].
 \label{app:axionkernels}
\end{align}
The source quotient $H_S'/x$ has the regular differentiated limit
$h_S'$ at zero transfer. The complete trace includes this information
before its heavy-target organization.

\subsection{Tensor source kernels before radial reduction}
\label{app:directtensor}
Use the mixed tensor matrix $E^\mu{}_{\nu}=\eta^{\mu\rho}_{\rm bulk}E_{\rho\nu}$.
The tensor vertices can be written using the densities
\begin{align}
 \ell_{qEE}&=n\mathcal R_2[E]+\tfrac n4\tr(E'^2)-C\tr(EE')
 -\tfrac12\partial^\rho\chi\tr(E'\partial_\rho E)
 +\tfrac12\tr(\mathsf C\{E,E'\}),\notag\\
 \ell_{EEE}&=\mathcal C_3^\Gamma[E]+\tfrac12\tr(EE'E'),\qquad
 C=\Box\chi,\quad \mathsf C^\mu{}_{\nu}=\partial^\mu\partial_\nu\chi.
 \label{app:TTvertices}
\end{align}
The complete direct ADM densities are
\begin{align}
 \ell_{qEE}^{\rm ADM}&=\ell_{qEE}+3A'n\tr(EE')
 +\tfrac14(\bar\phi'q'+e^{2A}\mathcal V_{,\phi}q)\tr(E^2),\notag\\
 \ell_{EEE}^{\rm ADM}&=\ell_{EEE}+3A'\tr(E^2E')
 +\left(2A'^2-\frac{\bar\phi'^2}{12}
              -\frac{e^{2A}\mathcal V}{6}\right)\tr(E^3).
 \label{eq:tensorunreduced}
\end{align}
Their differences from Eq.~\eqref{app:TTvertices} are
\begin{align}
 P(\ell_{qEE}^{\rm ADM}-\ell_{qEE})
  &=\tfrac14\partial_\xi[P\bar\phi'q\tr(E^2)],\notag\\
 P(\ell_{EEE}^{\rm ADM}-\ell_{EEE})
  &=\partial_\xi[PA'\tr(E^3)].
 \label{eq:surfaceidentities}
\end{align}
The full double-pole residues of these surfaces vanish on the stated
normalizable domains. Their cancellation is included before the
on-shell heavy-target organization.

\subsection{Polarized contraction and canonical-spin average}
\label{sec:polarizedtensor}\label{app:spinprojection}
Let the incoming momenta be $k_o=-p'$, $k_s=p'-p$ and $k_i=p$.
For lower-index external polarizations $a,b$ define
$\mathsf A=\eta a$, $\mathsf B=\eta b$ and
\begin{equation}
 S=\tr(\mathsf A\mathsf B),\quad
 L=k_s^{\rm T}\{\mathsf A,\mathsf B\}\eta k_s,\quad
 T=\tr[(\eta h)\{\mathsf A,\mathsf B\}].
 \label{eq:polarcontractions}
\end{equation}
Here $a,b$ label polarizations only. Let $R$ and $G$ be the bilinear
and trilinear coefficients of $\mathcal R_2$ and $\mathcal C_3^\Gamma$
with all field assignments included. The full response is expressed through
\begin{align}
 I_1&=\int\dd\xi\,Pu^2H_T,\quad
 I_2=\int\dd\xi\,P(\tfrac12u'^2H_T+uu'H_T'),\notag\\
 J_1&=-\int\dd\xi\,Pc^2H_Su^2/6,\quad
 J_2=-\int\dd\xi\,Pc^2H_Su'^2/12,\quad
 J_3=\int\dd\xi\,Pc^2H_S'uu'/(6x),\notag\\
 \rho_{oi}&=RJ_1+SJ_2+(3xS/2-L)J_3,\qquad
 \mathcal T_{ohi}=2(GI_1+TI_2).
 \label{app:TTradials}
\end{align}
These integrated full expressions include the vanishing full surfaces.
For a general metric insertion use its transverse-traceless part plus
$\Pi^{\mu\nu}h_{\mu\nu}\rho_{oi}/3$, where
$\Pi_{\mu\nu}=\eta_{\mu\nu}-k_{s\mu}k_{s\nu}/x$.
The physical trace is $-\rho_{oi}$.

Choose the five orthonormal spatial rest tensors
$(xx-yy)/\sqrt2$, $(xx+yy-2zz)/\sqrt6$, $(xy+yx)/\sqrt2$,
$(xz+zx)/\sqrt2$, and $(yz+zy)/\sqrt2$. Boost them without rotation to
$p=(E,0,0,-Q/2)$ and $p'=(E,0,0,Q/2)$, retaining the same rest-spin label
on both legs. With $h=\operatorname{diag}(1,1/2,1/2,0)$ and $y=x/\mu$,
the Bose-summed averages are
\begin{equation}
\begin{aligned}
 S&=(y^2+10y+30)/30,&L/\mu&=y(y+4)(y+5)/30,\\
 T&=-(y^2+2y-20)/30,&R/\mu&=-(y^3+10y^2+30y-60)/120,\\
 G/\mu&=-(y^3+16y^2+2y-40)/240.
\end{aligned}
 \label{app:spinsums}
\end{equation}
In particular $S(0)=1$, $T(0)=2/3$, $R(0)=\mu/2$ and $G(0)=\mu/6$.
These coefficients enter the tensor-source and trace contractions.

Set $U=PA'u^2$, $V_s=P\bar\phi'cu^2$. The regular unreduced kernels,
including the local terms in Eq.~\eqref{eq:tensorunreduced}, are
\begin{align}
 \overline A_{T,\rm full}&=\int\dd\xi\,(\alpha_AH_T+\beta_AH_T'),\notag\\
 \alpha_A&=\frac{GPu^2+(T/2)Pu'^2+3TU'}{\mu(1+y/4)},\qquad
 \beta_A=\frac{TPuu'+3TU}{\mu(1+y/4)},\notag\\
 \rho_{T,\rm full}&=\int\dd\xi\,(\alpha_\rho H_S+\beta_\rho H_S'),\notag\\
 \alpha_\rho&=-\frac R6Pc^2u^2-\frac S{12}Pc^2u'^2+\frac S2V_s',\notag\\
 \beta_\rho&=\frac16Pc^2uu'\left(\frac32S-\frac Lx\right)+\frac S2V_s,
 \qquad \overline A_{T,\rm full}^{\rm S}=-\rho_{T,\rm full}/(2\mu).
 \label{app:regularTensor}
\end{align}
The factor $L/x=(y+4)(y+5)/30$ is regular. Thus the contracted shift
quotient is regular in the full contracted source response. The radial
identities below recover the complete form factors in Eq.~\eqref{eq:maintensorfull}.

For conversion from any invariant spin-two basis~\cite{cotogno_multipoles},
form $\overline{\mathcal T}^{\mu\nu}=\tfrac15\sum_s\mathcal T_{ss}^{\mu\nu}$
with the same boosts. In the Breit frame with transfer along the third axis,
\begin{equation}
 \overline D_T=-\frac{2\overline{\mathcal T}^{11}}{Q^2},\qquad
 \overline A_T=\frac{\overline{\mathcal T}^{00}+\overline{\mathcal T}^{11}}{2E_Q^2},\qquad
 \overline A_T^{\rm S}=\frac{\overline{\mathcal T}^{00}-2\overline{\mathcal T}^{11}}{2M_T^2}.
 \label{app:spinmonopolemap}
\end{equation}
The calculation determines these three canonical-spin responses. The
complete eleven-invariant decomposition requires additional polarized
projections.

\subsection{On-shell tensor identities}
\label{app:projectionorder}
Define $R_u=\int Pu'^2H_T$ and $B_u=\int Puu'H_T'$. With the mode
and source equations and vanishing endpoint terms,
\begin{equation}
 B_u=-\frac{x}{2}I_1,\qquad R_u=(\mu+x/2)I_1,\qquad
 I_2=(\mu/2-x/4)I_1,\qquad J_3=J_1/2.
\label{app:I2identity}
\end{equation}
Substitution in the polarized response gives Eq.~\eqref{eq:maintensorfull}.
For the normalized trace overlap, $c^2=6[1+(1/A')']$ and
\begin{equation}
 \left[\frac{P}{A'}(u'^2+\mu u^2)\right]'
 =P\{[(1/A')'-3]u'^2+\mu[(1/A')'+3]u^2\}
\label{app:tensortraceNorm}
\end{equation}
imply $\int Pc^2(\mu u^2+u'^2)\dd\xi=24\mu$ and
$\mathcal G_T(0)=1$. The HT scalar-channel weight is obtained from the
on-shell organization of this complete expression, followed by the
finite-mass normalization in Eq.~\eqref{eq:HTweights}.

\section{HT weights, forward identities and positivity}
\label{app:forward}\label{app:HTderivation}
\subsection{Integrated target identity and the three leading weights}
For $-(Wu')'+WUu=\mu Wu$, integration of $(Wfuu')'$ and then of
$Wf'uu'$ gives Eq.~\eqref{eq:HTmodeidentity}. The endpoint terms vanish
for the normalizable target modes and regular probes. The identity keeps
the radial kinetic contribution at the same leading order as $\mu u^2$.
The large-eigenvalue organization is at fixed $x$ and fixed probe operators;
its remainder is expressed through background/probe derivatives. The normalization on the physical ground states is then fixed by
Eq.~\eqref{eq:HTweights}.

For the active scalar, isolate the leading target-kinetic terms of
Eq.~\eqref{app:scalarpolynomials}:
\begin{equation}
 F_0=-\frac{c^2}{6}(\mu_Su_S^2+u_S'^2)
 +F_{0,\rm background}.
\label{app:HTscalarleading}
\end{equation}
The tensor-source leading term is $\int Pu_S^2H_T$. In the normalized
trace, the two kinetic contributions combine using
Eq.~\eqref{eq:HTmodeidentity} with $f=c^2H_S$:
\begin{equation}
 \begin{aligned}
 &\frac1{12\mu_S}\int Pc^2(\mu_Su_S^2+u_S'^2)H_S\dd\xi
 =\frac16\int Pc^2u_S^2H_S\dd\xi\\
 &\hspace{10mm}+\frac1{12\mu_S}\int u_S^2
 \left[\frac12(P(c^2H_S)')'-Pc^2H_SU_S\right]\dd\xi.
 \end{aligned}
\label{app:HTscalaridentity}
\end{equation}
The remaining terms of $F_0,F_1,F_2,G_0,G_1$ are retained in the full
calculation and organized in the inverse-eigenvalue remainder; they are
not set to zero in the original action.

For the pseudoscalar define $g=cZ_{,\phi}/Z$, $a_c=c^2/6$, and $W=PZ$.
Its zero-explicit-transfer coefficient is
\begin{equation}
 -\frac{P\mathcal E_0}{2\mu_a}
 =\frac{W}{2\mu_a}
 [(a_c+g)u_a'^2+\mu_a(a_c-g)u_a^2].
\label{app:HTaxionkinetic}
\end{equation}
Using the target identity gives the leading overlap
$\int Wa_cu_a^2H_S\dd\xi$. The explicit $g$ contributions cancel
between the radial and boundary kinetic terms. Subleading source-gradient
and background derivatives retain additional $Z_{,\phi}$ dependence.
The leading norm density still depends on the axion coupling through $W$
and the normalizable mode.

For the tensor, the radial identity is exact:
\begin{equation}
 \mathcal G_T(x)=\frac1{12}\int Pc^2u_T^2H_S\dd\xi
 +\frac1{48\mu_T}\int u_T^2[P(c^2H_S)']'\dd\xi.
\label{app:HTtensoridentity}
\end{equation}
Together with the leading coefficients of the recoil polynomials, this
produces $\widetilde\rho^{(0)}_{S,T}=Pc^2u_T^2/12$ and
$\rho^{(0)}_{A,T}=Pu_T^2/2$. All three channels therefore give
$\widetilde\rho^{(0)}_{S,X}=c^2p_X/6$. The finite-mass model preserves
this shape and divides by its computed integral.

\subsection{Positive radial resolvent and complete monotonicity}
\label{app:HTpositivityProof}
For either metric-source channel, set
\begin{equation}
 L_Y=-\frac1{w_Y}\partial_\xi(w_Y\partial_\xi),\qquad
 (L_Y+x)H_Y=0,\qquad (w_T,w_S)=(P,Pc^2).
 \label{app:positiveSourceOperator}
\end{equation}
Regulate the radial interval by $\epsilon\leq\xi\leq R$, with
$H_Y(\epsilon,x)=1$ and vanishing infrared flux. Assume $w_Y>0$ on
this interval. The inverse of $L_Y+x$ on homogeneous ultraviolet data
and the same infrared condition is positivity preserving for $x\geq0$.
For example, multiplying $(L_Y+x)v=f\geq0$ by the negative part of $v$
and integrating with measure $w_Y\dd\xi$ excludes a negative solution.
The Dirichlet ultraviolet condition removes the constant zero mode.

The maximum principle gives $H_Y>0$. Since
$(L_Y+x)(1-H_Y)=x\geq0$ and $1-H_Y$ has homogeneous boundary data,
$H_Y\leq1$. Define $U_{Y,n}=(-1)^n\partial_x^nH_Y$.
The source boundary value is independent of $x$, so $U_{Y,n}$ has
homogeneous boundary conditions for every $n\geq1$, and
\begin{equation}
 (L_Y+x)U_{Y,n}=nU_{Y,n-1},\qquad U_{Y,0}=H_Y>0.
 \label{app:CMrecursion}
\end{equation}
Positivity of the inverse therefore implies $U_{Y,n}\geq0$ for every
$n$. Removing the regulators along the regular source branch preserves
the pointwise bounds. A finite pointwise limit of completely monotone
functions is completely monotone on $x>0$; equivalently, each regulated
source is a Laplace transform of a positive measure, and this property
survives the regular limit. Boundary derivatives at $x=0$ are understood
only when their moments are finite.

Integrating against either of the normalized positive measures
$p_X\dd\xi$ and $c^2p_X\dd\xi/\langle c^2\rangle_X$ preserves complete
monotonicity and the unit forward value. This proves
Eqs.~\eqref{eq:HTpositivity}--\eqref{eq:HTcompleteMonotonicity}.
The proof uses only the normalized positive weights and the regular
source domain of the model. The physical trace is precisely $\mathcal S_X$ and obeys the same
complete-monotonicity inequalities. No sign for $D$ or local Breit
densities follows.

\subsection{Exact finite-mass trace and regular forward limit}
Covariantly normalized states and the conserved total EMT give $A(0)=1$
and $\Theta(0)=2M^2$. Contracting Eq.~\eqref{app:EMT} yields
\begin{equation}
 \Theta(t)=(2M^2-t/2)A(t)-3tD(t)/2,\qquad
 D(-Q^2)=\frac{4M^2}{3Q^2}(A^{\rm S}-A)-\frac13A.
\label{app:traceidentityforward}
\end{equation}
The HT pair has $A^{\rm S}_{\HT}=\mathcal S$ and $A_{\HT}=\mathcal A$,
so the exact identity retains $-\mathcal A/3$ in $D$.
Its finite derivative is
\begin{equation}
 D_{\HT}(0)=\frac{4\mu}{3}(\dot{\mathcal S}(0)-\dot{\mathcal A}(0))-\frac13,\qquad
 \dot A^{\rm S}_{\HT}(0)=\dot{\mathcal S}(0).
\label{app:HTDderivative}
\end{equation}
Without the normalization of $\mathcal S$, the bare ground-state overlap
would introduce $4\mu(\langle c^2\rangle/6-1)/(3x)$ into $D$.
The HT normalization removes this pole and fixes a regular forward stress coefficient.

The normalizations and differentiated forward identities were checked
numerically for all three reference targets.
\FloatBarrier

\section{Spectra-and-Wilson calibration}
\label{sec:calibration}\label{app:numerics}
The six coefficients in Table~\ref{tab:mainparameters} are determined from
six continuum mass ratios $M/\sqrt\sigma$ (two in each target channel)
from Ref.~\cite{at}, ten static-force values from Ref.~\cite{necco}, and
$r_0\sqrt\sigma=1.160(6)$~\cite{at}. The common Sommer convention is
$r_0=0.49\,\fm$, with $r_0^2F(r_0)=1.65$ fixing $m$. The objective is
\begin{equation}
 \chi^2_{\rm cal}=\chi^2_{\rm spec}+\chi^2_{\rm force}
                   +\chi^2_{r_0\sqrt\sigma}.
 \label{eq:mainchi2}
\end{equation}
Neither a GFF, its covariance, nor a derived radius enters this fit.
The spectral and force blocks use diagonal working covariances from the
published errors; correlations within and between these blocks are not
available. With 17 observations and six fitted coordinates, the nominal
calibration count is $\nu_{\rm cal}=11$.

\begin{table}[!htbp]\centering\small
\begin{tabular}{lrrr}\toprule
Comparison & $\chi^2$ & $\nu$ & $\chi^2/\nu$\\\midrule
Calibration: spectra & 6.28 & \textemdash & \textemdash\\
Calibration: static force & 7.15 & \textemdash & \textemdash\\
Calibration: $r_0\sqrt\sigma$ & 13.40 & \textemdash & \textemdash\\
Calibration total & 26.83 & 11 & 2.44\\\midrule
Scalar GFF prediction: Full & 73.60 & 27 & 2.73\\
Scalar GFF prediction: HT & 16.98 & 27 & 0.63\\\bottomrule
\end{tabular}
\caption{Calibration residuals and subsequent GFF comparisons at common
parameters. The calibration components share six fitted coordinates;
only their total is assigned 11 degrees of freedom. Each GFF row uses
27 correlated observations with no GFF parameter adjustment.}
\label{tab:mainresiduals}
\end{table}

The string ratio is $r_0\sqrt\sigma=1.18196$, which differs from
$1.160(6)$ by $3.66$ input standard errors and supplies $13.40$ to the
objective. This residual belongs to the effective-string sector and is
common to Full and HT. The fitted region constrains the dimensionless
quartic warp-deformation coefficient
$\omega_4\equiv(y_{\rm IR}-y_{\rm UV})/b_c^2=0.008570$ through
$F=-y_{\rm UV}b^2/2-\omega_4 b^4/4+O(b^6)$ more directly than the
ultimate infrared radial slope $s_\infty$.

\paragraph{Gravitational-form-factor comparison.}
After calibration, the predictions are evaluated at fourteen scalar $A$
and thirteen nonzero-transfer $D$ momenta~\cite{gff,release} using their
complete $27\times27$ covariance:
\begin{equation}
 \chi^2_{AD}=(\mathbf f_{AD}-\mathbf d_{AD})^{\rm T}
 C_{AD}^{-1}(\mathbf f_{AD}-\mathbf d_{AD}).
 \label{eq:predictionchi2}
\end{equation}
At the fixed parameter point, $\nu_{AD}=27$ and no background coefficients
are subtracted from this prediction count. With the nominal fixed
covariances, the upper-tail probabilities are $3.39\times10^{-6}$ for
Full and $0.9316$ for HT; the corresponding calibration value is $0.00487$
for $\nu_{\rm cal}=11$. The HT lower-tail probability is $0.0684$.
These are conditional residual diagnostics at the fixed background and
working covariances, rather than model-selection probabilities.

For context, joint $n$-pole fits to the same final renormalized data vector
and full covariance use
$A_n=(1+b_AQ^2)^{-n}$ and $D_n=D_0(1+b_DQ^2)^{-n}$, with
$A_n(0)=1$. The three fitted parameters $b_A,b_D,D_0$ give
$\chi^2=14.42,12.67,12.78$ for $n=1,2,3$, respectively, with 24 nominal
degrees of freedom. These benchmark fits use the final means and
covariance, rather than reproduce the irrep-by-irrep normalization and
resampling procedure of Ref.~\cite{gff}. Reduced residuals near $0.5$--$0.6$
therefore also occur for fitted conventional parametrizations of these
data. The lattice calculation uses one spacing; the authors caution that
the nearly flat $D(t)$ has limited statistical precision and report a
statistically compatible three-stout analysis~\cite{gff}.

Profiling a common multiplicative factor $g$ on the lattice prediction
vector $(A,D)$, with the covariance fixed, gives $g_{\HT}=0.99988$ and
changes $\chi^2_{AD}$ by less than $0.001$. This is an observation-scale
sensitivity; the physical form factors remain normalized by $A(0)=1$.

\paragraph{Relative physical-scale matching.}
\label{app:scaleMatching}
The model uses $r_0=0.49\,\fm$, while the GFF points retain their published
physical momentum axis. The lattice study quotes $\beta=5.95$ and
$a=0.098\,\fm$~\cite{gff,abbott2024}. At this coupling, Table~1 of
Ref.~\cite{necco} gives $r_0/a=4.898(12)$, whereas its interpolation gives
$4.91223$; the rational interpolation of Ref.~\cite{durr2007} gives
$4.91124$. Combined with the printed spacing, these imply
$r_0\simeq0.4800$--$0.4814\,\fm$. The rounded spacing and public GFF record
do not specify a unique exact conversion for the released momentum
array~\cite{gff,release}. Ref.~\cite{durr2007} itself uses $r_0=0.49\,\fm$
with a four-percent pure-Yang--Mills unit-conversion ambiguity, which is
not a separately quoted scale error of the GFF ensemble.

For a lattice length assignment $R_{\rm lat}$ and model assignment
$R_{\rm mod}=0.49\,\fm$, consistent reexpression gives
\[
 Q^2_{\rm common}=\left(\frac{R_{\rm lat}}{R_{\rm mod}}\right)^2
 Q^2_{\rm published},\qquad
 r_{\rm common}=\frac{R_{\rm mod}}{R_{\rm lat}}r_{\rm published}.
\]
The inferred assignments give momentum factors $0.9596$--$0.9652$ and
convert the published scalar radius to approximately $0.268(32)\,\fm$.
We keep the published axis for the primary comparison and assess the
relative matching by $m\to(1\pm0.02)m$ at fixed dimensionless dynamics.
The resulting $\chi^2_{AD}$ ranges are $69.69$--$77.55$ for Full and
$16.05$--$18.00$ for HT. These are sensitivity ranges, not scale-confidence
intervals. The Necco--Sommer interpolation-based axis conversion gives
$77.13$ and $17.89$, respectively. A common change of units leaves the
dimensionless string-ratio discrepancy unchanged.

\paragraph{Conditional parameter uncertainties.}
The covariance of the six calibrated coordinates is obtained from the
inverse Gauss--Newton matrix of the whitened 17-component residual,
without multiplication by $\chi^2/\nu$. Propagation through the mode and
source problems gives Table~\ref{tab:HTuncertainties}. The derivatives
were checked with two step sizes. These errors condition on the chosen
functional forms, fixed $b_a=3$, the infrared-decaying axion branch, the
model scale, and the diagonal working spectral/force covariances.
Relative lattice-scale matching, discrete branches, functional variations
and continuum extrapolation are outside these local errors. The string-ratio
residual does not establish a common multiplicative underestimate of all
input variances, so no uniform error inflation is applied.

\begin{table}[!htbp]\centering\small
\begin{tabular}{lrrr}\toprule
Target & $D(0)$ & $r_A$ [fm] & $r_{\rm mass,BF}$ [fm]\\\midrule
$0^{++}$ & $-1.3706\pm0.0059$ & $0.2394\pm0.0027$ & $0.2762\pm0.0023$\\
$0^{-+}$ & $-2.6852\pm0.0529$ & $0.2640\pm0.0091$ & $0.2999\pm0.0079$\\
$2^{++}$ (canonical spin) & $-2.3078\pm0.0350$ & $0.2776\pm0.0015$ & $0.3112\pm0.0013$\\\bottomrule
\end{tabular}
\caption{HT predictions with local linearized uncertainties propagated
from the spectra-and-Wilson calibration. The errors are not multiplied by
$\sqrt{\chi^2_{\rm cal}/\nu_{\rm cal}}$ and condition on diagonal working
spectral/force covariances and the reference axion branch. Branch,
parametrization and relative-scale sensitivities are separate.}
\label{tab:HTuncertainties}
\end{table}

\paragraph{Focused parametrization checks.}
\label{app:focusedParametrization}
We varied $b_a$ over seven values spanning $1.5$--$6$ and rematched
$\zeta_1,\zeta_2$ to the same two pseudoscalar masses, with the other
reference quantities fixed. On the infrared-decreasing $Z$ branches, the HT radius
$r_{A,a}$ ranges from $0.237$ to $0.268\,\fm$ and
$r_{{\rm mass,BF},a}$ from $0.278$ to $0.303\,\fm$, while
$D_a(0)$ remains between $-2.737$ and $-2.676$.
Infrared-growing solutions were followed at $b_a=1.5,2,2.5,3,3.5,4$;
they give $r_{{\rm mass,BF},a}=0.332$--$0.393\,\fm$ and
$D_{a,\HT}(0)=-2.582$ to $-2.488$. At the reference $b_a=3$, the
additional solution $(\zeta_1,\zeta_2)=(13.5733,-32.3562)$ gives
$M_{a,0}=2.5113\,\GeV$, $M_{a,1}=3.4699\,\GeV$ and a third level
$4.2540\,\GeV$, compared with $4.2976\,\GeV$ on the reference branch.
Its positive $Z$ reaches approximately $9.19\times10^{-4}$ before
rising in the infrared. Neither $Z>0$ nor $\log Z=O(\log b)$ excludes it.
The branch changes the reference mass radius to $0.3693\,\fm$ and the
HT $D$-term to $-2.4997$; Full gives $-2.4863$ there. Thus $D_a(0)$ is
less sensitive in relative terms than the radius, but its sampled
between-branch spread still exceeds the local error in
Table~\ref{tab:HTuncertainties}. The ranges are sampled model sensitivities,
not confidence intervals or an exhaustive branch classification. The
reference $\zeta_1<0$ choice is held fixed in all primary curves.

\phantomsection\label{app:warpInterpolation}
\emph{Warp-profile interpolation.}
Equation~\eqref{eq:warpDerivativeProfile} relates $Y$ directly to the
warp deformation. For the alternative exponential interpolation,
\begin{equation}
\begin{aligned}
 Y_{\exp}(b)&=y_{\rm UV}+(y_{\rm IR}-y_{\rm UV})
                     [1-e^{-b^2/b_c^2}],\\
 F_{\exp}(b)&=-\frac{y_{\rm IR}}2b^2
 +\frac{(y_{\rm IR}-y_{\rm UV})b_c^2}{2}
                     [1-e^{-b^2/b_c^2}].
\end{aligned}
\label{app:exponentialWarp}
\end{equation}
We reconstruct $K,V,c$ consistently from this profile and fit only its
two warp coefficients to the same four scalar/tensor lattice mass
observations, holding the Wilson sector and dimensional scale fixed.
The resulting coefficients are $y_{\rm IR}^{\exp}=0.472460$ and
$b_c^{\exp}=6.97997$, with spectral $\chi^2=6.18$ compared with $6.28$
for the reference rational profile.

Both interpolation families have the small-$b$ expansion
\begin{equation}
 F(b)=-\frac{y_{\rm UV}}2b^2-\frac{\omega_4}{4}b^4+O(b^6),
 \qquad \omega_4\equiv\frac{y_{\rm IR}-y_{\rm UV}}{b_c^2}.
 \label{app:quarticWarpCoefficient}
\end{equation}
For equal parameters their Taylor series first differ at order $b^6$.
The fitted quartic deformation is nearly unchanged:
$\omega_{4,\rm rat}=0.008570$ and $\omega_{4,\exp}=0.008557$.
In contrast, the asymptotic radial mass-squared slope defined in
Eq.~\eqref{eq:mainreggeslope} changes from
$s_{\infty,\rm rat}=16.56\,\GeV^2$ to
$s_{\infty,\exp}=9.32\,\GeV^2$.
The scalar/tensor HT forward $D$-terms change by less than $0.005\%$
and their radii by less than $0.03\%$. These low-lying overlap observables
are therefore stable under the tested change of distant-infrared
completion, despite the different asymptotic radial spectra.

The scalar finite-transfer response is similarly stable. Evaluating both
HT predictions at the same 27 lattice $A,D$ observations with the complete
covariance gives
\begin{equation}
 \left.\frac{\chi^2_{AD}}{27}\right|_{\rm rat}
       =\frac{16.9814}{27}=0.6289,\qquad
 \left.\frac{\chi^2_{AD}}{27}\right|_{\exp}
       =\frac{16.9931}{27}=0.6294.
 \label{app:warpGFFComparison}
\end{equation}
Neither calculation adjusts parameters to GFFs; the difference
$\Delta\chi^2_{AD}=0.0117$ is negligible for this comparison.
At large transfer, the common ultraviolet source equations and
$u_S\sim u_3\xi^3$ give, with $k=Q/m$,
\begin{equation}
 A_{S,\HT}(k^2)\sim\frac{48u_3^2}{k^4},\qquad
 A^{\rm S}_{S,\HT}(k^2)\sim
       \frac{384u_3^2}{\langle c^2\rangle_S k^6},\qquad
 D_{S,\HT}(k^2)\sim-\frac{16u_3^2}{k^4}.
 \label{app:warpScalarUV}
\end{equation}
The normalized mode coefficient $u_3$ and $\langle c^2\rangle_S$ are
computed separately for each background. The leading coefficients of
$A_S$ and $D_S$ differ by approximately $0.057\%$, and that of the
physical trace by $0.070\%$. At the sampled transfers
$Q^2=2.365,10,100\,\GeV^2$, the relative differences in all three
scalar HT form factors are below $0.08\%$.

The radii probe the small-transfer slopes of the complete source response,
whereas large spacelike transfer emphasizes the near-boundary region.
Their simultaneous stability shows that the tested change of
distant-infrared completion has little effect on these overlap
observables. In a spectral representation the source response sums radial
poles together with their residues; the changed asymptotic masses
therefore need not produce a comparable change in the resummed GFF.

The normalizable target and regular source equations, radial overlaps,
and Fourier integrals were evaluated numerically. Source normalization,
the finite-mass EMT identity and the pressure/shear moments were checked
against the independently differentiated form factors. The final curves
and their fixed parameters are provided in the accompanying data tables.
\FloatBarrier

\end{document}